\documentclass[]{aa}
\usepackage[varg]{txfonts}
\usepackage{graphicx}
\usepackage{booktabs}
\usepackage{amsmath}
\usepackage{colortbl}
\usepackage{subfig}
\usepackage{rotating}
\usepackage{amssymb}
\usepackage{latexsym}
\usepackage{ifthen}
\usepackage{enumitem}

\begin{document}

\title{An X-ray/SDSS sample (I): multi-phase outflow incidence and dependence on AGN luminosity} 

\author{M. Perna
		\inst{\ref{i1},\ref{i2}}\thanks{E-mail: michele.perna4@unibo.it}
		\and 
	G. Lanzuisi
		\inst{\ref{i1},\ref{i2}}	
		\and
	M. Brusa
		\inst{\ref{i1},\ref{i2}} 
		\and
	M. Mignoli
		\inst{\ref{i2}}
	\and 
	G. Cresci
		\inst{\ref{i4}} 
}


\institute{Dipartimento di Fisica e Astronomia, Universit\`a di Bologna, viale Berti Pichat 6/2, 40127 Bologna, Italy\label{i1}
	\and
	INAF - Osservatorio Astronomico di Bologna, via Ranzani 1, 40127 Bologna, Italy\label{i2}
	\and
	INAF - Osservatorio Astrofisico di Arcetri, Largo Enrico Fermi 5, 50125 Firenze, Italy\label{i4}
}

\date{Received 2 November 1992 / Accepted 7 January 1993}

\abstract {} 
{
The connection between the growth of super-massive black holes (SMBHs) and the
evolution of their host galaxies is nowadays well established, though the underlying mechanisms explaining their mutual relation are still debated. Multi-phase fast, massive outflows have been postulated to play a crucial role in this process. The aim of this work is to constrain the nature and the fraction of outflowing gas in active galactic nuclei (AGNs), as well as the nuclear conditions possibly at the origin of such phenomena. 
} 
{
We present a large spectroscopic sample of X-ray detected SDSS AGNs at z $<0.8$ having a high signal-to-noise ratio in [O {\small III}]$\lambda$5007 line, in order to unveil the faint wings of the emission profile associated with AGN-driven outflows. X-ray and optical flux ratio diagnostics are used to select the sample. Physical and kinematic characterization have been derived re-analysing optical (and X-ray) spectra.
} 
{
We derive the incidence of ionised ($\sim40\%$) and atomic ($< 1\%$) outflows covering a wide range of AGN bolometric luminosity, from $10^{42}$ to $10^{46}$ erg/s. We also derive bolometric luminosities and X-ray bolometric corrections to test whether the presence of outflows is associated with an X-ray loudness, as suggested by our recent results obtained studying high-z QSOs.
} 
{
We study the relations between the outflow velocity inferred from [O {\small III}] kinematic analysis and different AGN power tracers, such as black hole mass (M$_{BH}$), [O {\small III}] and X-ray luminosity. We show a well defined positive trend between outflow velocity and $L_X$, for the first time over a range of 5 order of magnitudes. 
Overall, we find that in the QSO-luminosity regime and at M$_{BH}>10^8$ M$_\odot$ the fraction of AGNs with outflows becomes $>50\%$. 
Finally, we discuss our results about X-ray bolometric corrections and outflow incidence in cold and ionised phases in the context of an evolutionary sequence allowing two distinct stages for the feedback phase: an initial stage  characterized by X-ray/optical obscured AGNs in which the atomic gas is still present in the ISM and the outflow processes involve all the gas components, and a later stage associated with unobscured AGNs, which line of sight has been cleaned and the cold components have been heated or exhausted. 
} 

\maketitle
\titlerunning{X-ray loudness and outflows} 

\section[Introduction]{Introduction}

Feedback mechanisms during the bright phase of Active Galactic Nuclei (AGN) 
are indicated as the leading processes responsible for the joint evolution of super-massive black holes (SMBHs) and galaxies (e.g., \citealt{Fabian2012,King2015}). 

Evolutionary models for the co-eval growth of galaxies and AGN predict for the quasar population (L$_{bol}>$10$^{45}$ erg/s) a relatively short ($<<$500Myr) three stages phase triggered by major merger events \citep{Menci2008,Hopkins2008}. In this evolutionary framework, the funnelling of large amount of gas into the nuclear region ignites a concomitant growth of the central SMBH and the host bulge through efficient star formation (SF), in a dust-enshrouded environment of dense gas. When the SMBH achieves a critical mass, establishing the M-$\sigma$ relation, it becomes powerful enough to affect the host galaxy. In the case of high accretion rates, the energy output of the SMBH couples to the different phases of the interstellar medium (ISM). During this coupling, galaxy-wide winds are expected to quench the galaxy SF, by reheating  the gas or pushing it out of the galactic potential well. 

The presence of AGN-driven outflows is now quite well established through high resolution observations of local and  high-redshift galaxies, at different wavelengths tracing the different phases (ionised,  neutral, and molecular) of the gas in the ISM \citep[e.g.,][]{Brusa2015,Cimatti2013,Feruglio2010,Glikman2012,Lanzuisi2015,Liu2013,Rupke2005b,Sturm2011,Talia2016,Villar2014}. 
Moreover, 
with the advent of high resolution, sensitive integral field spectrographs and millimetre interferometers, it is now possible to study in details the feedback phenomena, characterizing the galaxy-wide extension and the morphology of the ejected material as well as the masses and the energetics related to outflows \citep[e.g.,][]{Brusa2016,Carniani2015, Cicone2014,Cresci2015, Harrison2012, Harrison2014,Perna2015a,Perna2015b}. 
However, both the triggering feedback mechanisms and the physical processes responsible for the coupling between the AGN winds and the ISM resulting in outflows remain largely unknown. 
The full characterization of the AGN-host galaxy system is needed to discriminate between the details of various model realizations (see, e.g., the detailed discussion in \citealt{Brusa2015}): multiwavelength data are essential to derive nuclear and host properties associated with the presence of outflows, while synergies between major facilities (e.g., ALMA, NOEMA, SINFONI, MUSE) are required to study the multiphase wind/ISM interactions. In this respect, we note that spatially resolved measurements so far have been mostly limited to small/biased samples (see above references), and do not permit the exploration of a wide parameter space for the study of the feedback phenomena.

A different strategy is given by the analysis of large area optical surveys, such as Sloan Digital Sky Survey (SDSS). In recent times, a number of studies have focused on the analysis of the kinematics of the [O {\small III}]$\lambda$5007 line ([O {\small III}] hereinafter) with the main goal of inferring the presence of ionized outflows \citep[e.g.,][]{Bae2014,Concas2017,Komossa2008,Woo2016}. By combining SDSS with multi-wavelength data sets, several constraints on the properties and the effects of such outflows can be derived, at least in a statistical sense. For example, \citet{Mullaney2013}, combining SDSS data with radio luminosity ($L_{1.4 GHz}$), found a strong correlation between $L_{1.4 GHz}$ and the width of [O {\small III}], suggesting a connection between the presence of compact radio cores and the outflow phenomena (see also \citealt{Zakamska2016}; see \citealt{Woo2016} for a different interpretation). Other recent studies related the presence of outflows with star formation activity using several diagnostics from Herschel and Spitzer data \citep{Balmaverde2016,Wylezalek2016}, pointing to different conclusions regarding the role of AGN feedback (see also \citealt{Mullaney2015}). Despite some contradictory conclusions, all these works have shown that a large fraction of SDSS AGNs presents signatures of outflows in their ionized phase, from $\sim 20-40\%$ to $\sim$ $50-70\%$ depending on whether the objects are type 2 or type 1 AGNs \citep{Woo2016,VeronCetty2001}.
Moving to higher redshifts, \cite{Harrison2016} analysed a sample of $\sim$ 40 X-ray selected objects at z $\sim 1$ and reported a fraction of outflows $\sim 50\%$. Moreover, our recent works in the framework of
XMM-COSMOS survey \citep{Brusa2015,Brusa2016,Cresci2015,Perna2015a,Perna2015b} have shown that the few ($\sim$10 sources) luminous obscured AGN at z $\sim1.5$ with evidence of outflows collected so far, have relatively low X-ray $k_{bol}$ corrections (median L$_{bol}$ /L$_X$ $\approx$ 10). For comparison, more typical ratios of $\sim40$ are found for sources at the same median bolometric luminosity (\citealt{Lusso2012}). 
In order to validate a connection between outflow phenomena and X-ray
loudness, a systematic study of outflow signatures in X-ray selected samples is needed.  

This work is part of a series of papers investigating the physical and demographic characterization of the AGN-galaxy system during the feedback phase.
In this first paper we present a large sample of X-ray selected AGN at z $<0.8$, for which SDSS spectra are available.  Similarly to the analysis derived by previous works that combined optical spectroscopic analysis with radio and/or infrared wavelength bands, it extends the analysis for the first time to an X-ray selected large sample ($>100$  AGNs). 
We present the X-ray/SDSS sample selection procedure and the results obtained from relevant optical diagnostics. We focus on the incidence of outflows that can be derived studying ionized and atomic features in optical spectra, which is crucial to better constrain the temporal lengths of the different AGN phases in the context of evolutionary models (e.g., \citealt{Hopkins2008}). We also derive the X-ray bolometric corrections for each AGN and test the role of X-ray activity in the context of the feedback phase. 

The paper is organized as follows: in Section \ref{sdsssel} we outline our sample selection procedure; in Section \ref{sdssanalisi} we describe our multicomponent line fitting routine and the different emission line diagnostics. BPT diagrams are used to discriminate between star forming galaxies and AGNs among the X-ray selected sources. In section \ref{nuclearSDSS} we estimate their nuclear properties. In sections \ref{sdssincidence} and \ref{neutraloutflows} we derive the outflow fraction in the ionized and neutral phase, respectively, for different AGN sub-samples. In Section \ref{sdssloudness} we test if a X-ray loudness is actually associated with outflow processes. 
Finally, we summarize our results and their implications in the last section (Sect. \ref{sdssdiscussion}). A flat universe model with a Hubble constant of $H_0=$ 72 km s$^{-1}$ Mpc$^{-1}$, $\Omega_M$= 0.27 and $\Omega_{\lambda}$= 0.73 is adopted.

\section{Sample}\label{sdsssel}

\begin{figure}[t]
\centering
\includegraphics[width=9cm,trim=0 205 0 80,clip]{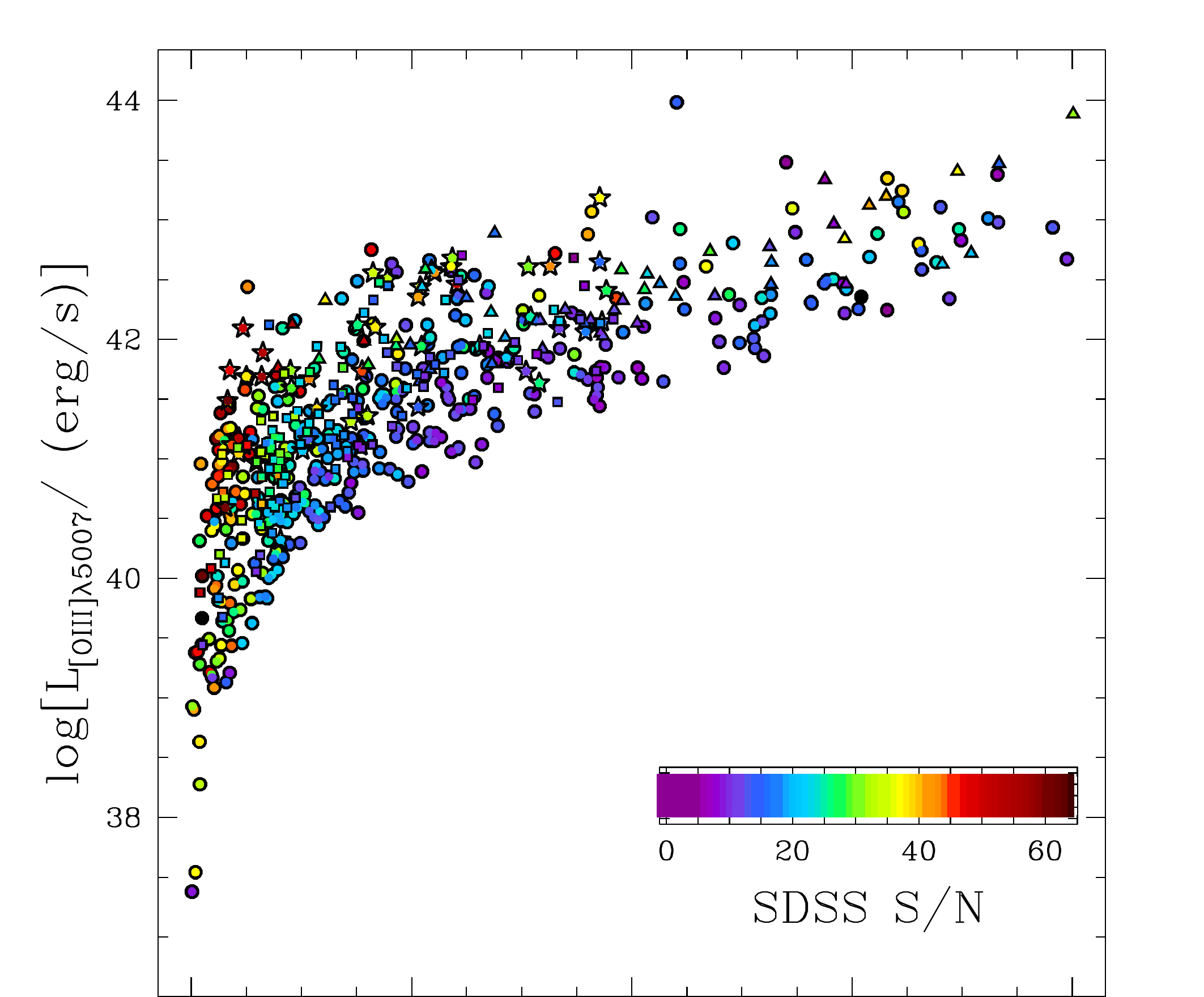}

\includegraphics[width=9cm,trim=0 130 0 100,clip]{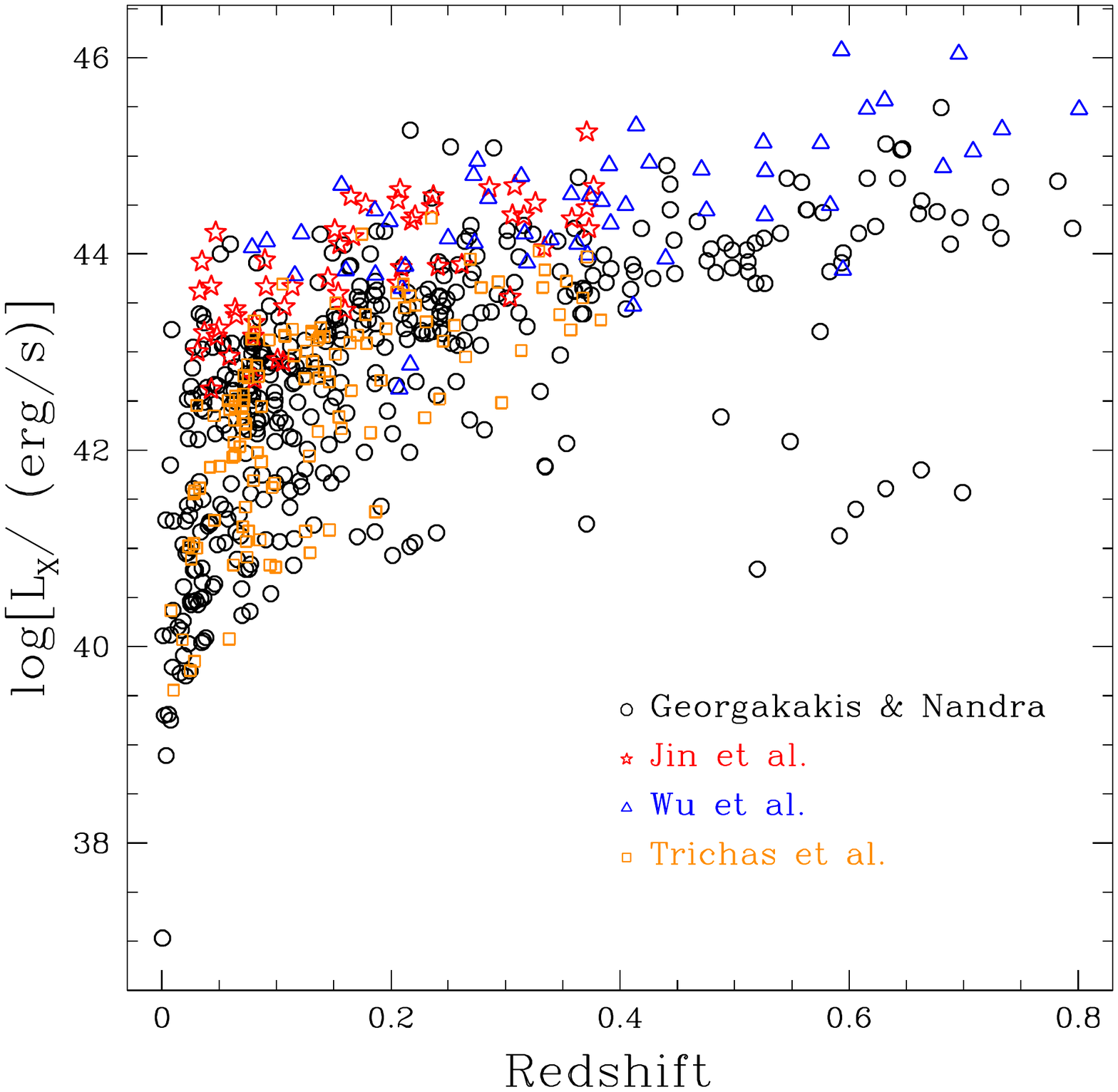}
\caption{({\it top panel:}) \small [O {\small III}]$\lambda$5007 total luminosity as obtained by our spectral analysis vs. redshift for the 624 candidate AGNs, coloured by nominal SDSS signal to noise ratio as labelled in the colour-bar. Circles, stars, triangles and squares refer to \cite{Georgakakis2011,Jin2012,Wu2012,Trichas2013} sub-samples respectively, used to construct our X-ray/SDSS sample. ({\it bottom panel:}) Absorption corrected X-ray luminosity vs. redshift fot the same sample of targets, distinguishing between the different sub-samples, as labelled in the figure. }
\label{zdistribution}
\end{figure}

We construct our AGN sample by analysing SDSS Data release 7 (DR7) spectra associated with X-ray emission taking advantage from previous studies in which the identification of the X-ray detected sources with optical counterparts has been performed through accurate methods. In particular, we consider: 
\begin{itemize}
\item
the catalog compiled by \citet{Georgakakis2011} obtained from the Serendipitous XMM Survey (\citealt{Watson2009}) in the area of the SDSS (XMM/SDSS), containing 2067 X-ray point sources with spectroscopic observations from the SDSS DR7 (\citealt{Abazajian2009}); 
\item
the \citet{Jin2012} sample of 51 unobscured type 1 AGNs selected to have high-quality spectra from both XMM-Newton and the SDSS DR7; 
\item
the \citet{Wu2012} catalog, whose 1034 objects are identified by matching the {\it Swift} pointings with SDSS DR5 quasar catalogue;
\item
the \citet{Trichas2013} catalogue, containing 617 matched sources from the Chandra Source Catalogue and the SDSS  DR7 spectroscopic sample (CSC/SDSS) at z$<$0.4.  
\end{itemize}

From this original sample of 3769 targets, we select the z $<0.8$ galaxies, to include in the SDSS wavelength coverage the [O {\small III}] emission line. 
We select all sources with S/N in the [O {\small III}] region $>10$, in order to {\it i)} unveil and analyse the faint wings of the [O {\small III}] profile, {\it ii)} to exclude line-less galaxies, which are typically red galaxies with no sign of AGN activity. After a final check for possible duplicates in the catalogues, we obtain a X-ray/SDSS sample containing 624 unique objects. 

The SDSS spectra are downloaded from the SDSS archive and re-analysed  using a multicomponent fitting routine detailed below.
Figure \ref{zdistribution}, top panel, shows the [O {\small III}] luminosity plotted against redshift, colour-coded according to increasing nominal SDSS S/N of each spectrum. 

X-ray properties are obtained from available spectral analysis results published by \citet{Jin2012}, \citet{Trichas2013} and \citet{Wu2012}. For the \citet{Georgakakis2011} 
 sample, only non-corrected luminosities were available. We therefore extracted X-ray spectra for all the sources in this sub-sample  from XMM archive, applying standard spectral fit procedures with a simple or double power-law  model to correct the luminosity for moderate absorption \citep{Lanzuisi2013}.

The X-ray sub-samples we used all come from heterogeneous
X-ray observations and the flux limits differ significantly from observation to observation. 
However, the final coverage of the X-ray luminosity versus redshift plane obtained from the X-ray/SDSS sample (Fig. \ref{zdistribution}, lower panel)
is similar to that of a multi-layer (''wedding cake'') survey, with a large-area,
shallow layer, corresponding to typical flux limit of  $f_{2-10\ keV}\sim 3\times 10^{-14}$ erg/cm$^2$/s, and a much smaller area covered
at deeper flux limits, down to $\sim 5\times 10^{-16}$ erg/cm$^2$/s.

\begin{figure}[t]
\centering
\includegraphics[width=9cm,trim=0 60 400 10,clip]{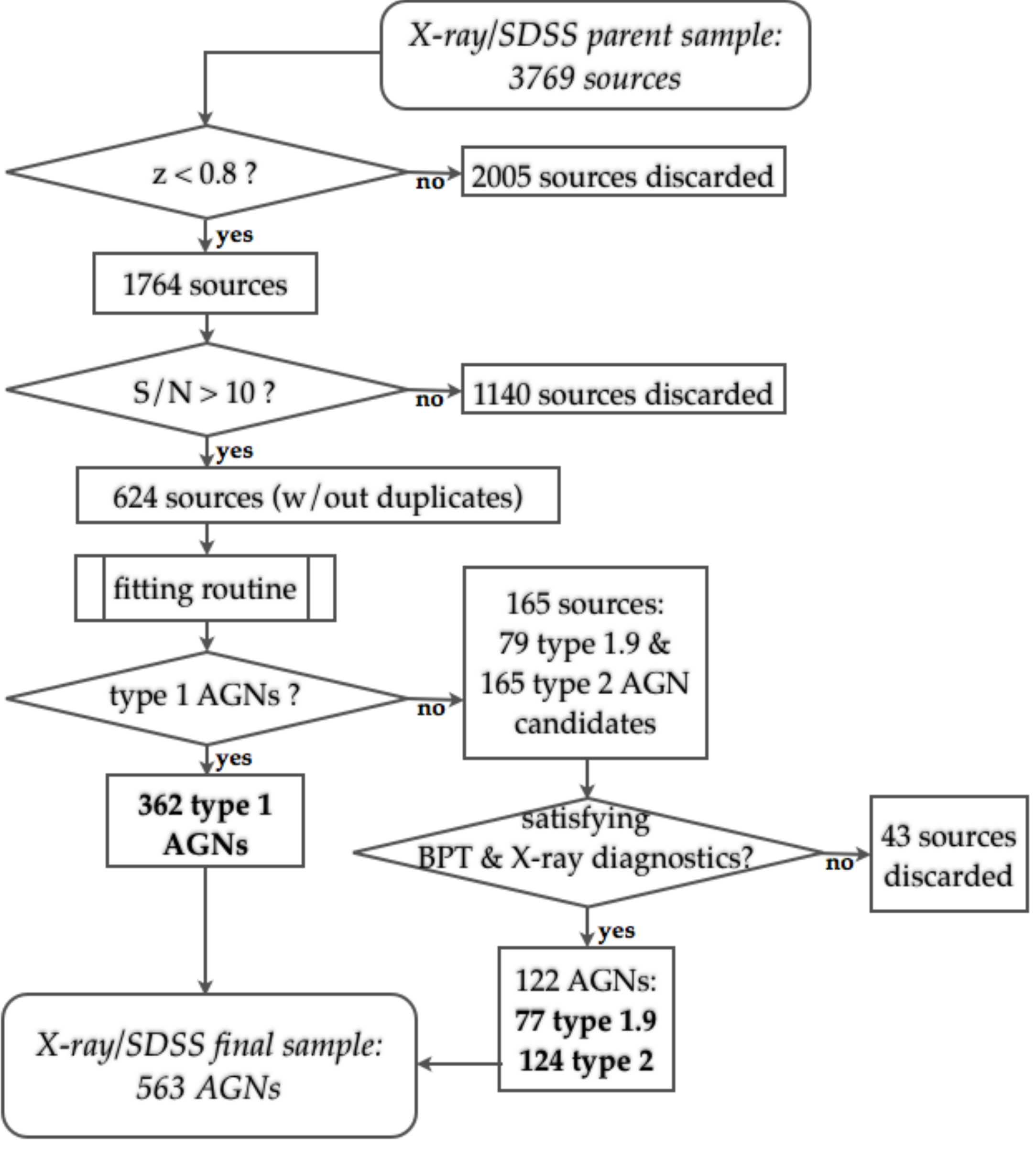}
\caption{\small Flow diagram representing the selection criteria (Sect. \ref{sdsssel}) and the diagnostics (Sect. \ref{sdssanalisi}) we used to define our X-ray SDSS sample.}
\label{flowdiagram}
\end{figure}

Figure \ref{flowdiagram} (top panel) shows a schematic diagram of our selection process, and summarizes the above mentioned criteria and the diagnostics displayed in the next sections.

\section{Spectral Analysis}\label{sdssanalisi}

\subsection{Multicomponent simultaneous spectral fit}

To characterize the nuclear accretion properties and determine kinematic and physical properties of the emitting gas, we proceed implementing the fitting prescriptions exposed in our previous studies \citep[e.g.,][]{Brusa2015} to characterize the permitted Fe {\small II} ($\lambda\lambda4000-5400\AA$) emission features and the He {\small II}$\lambda$4863 line in the case of typical type 1 spectra. 

We simultaneously fit each of the most prominent emission lines,
from the He {\small II} to [S {\small II}] doublet with four (at maximum) sets of Gaussian profiles:
\begin{itemize}
\item[BC]
-- one (two) Broad component(s): three Gaussian (or Lorentzian)
functions for the  He {\small II}, H$\beta$ and H$\alpha$ BLR emission
lines, with a FWHM greater than 1000 km/s. A second BC is added when
H$\alpha$ and H$\beta$ BLR cannot be modelled with a single broad
gaussian profile (see, e.g. \citealt{Shen2012});
\item[NC]
-- a Narrow component: nine Gaussian lines, one for each emission line
in the two regions of interest (i.e. He {\small II}, H$\beta$, [O {\small III}] doublet, H$\alpha$, [N {\small II}] and [S {\small II}] doublets) to account for the presence of unperturbed systemic emission (from the NLR or the host galaxy). The width of this kinematic component is set to be $\lesssim 550-600$ km/s;
\item[OC]
-- one (two, in an handful of cases) Outflow component(s): eight Gaussian lines, one for each emission line (with the exclusion of the He {\small II}, which is usually faint and never well constrained) to account for the presence of outflowing ionised gas. No upper limits are fixed for the width of this kinematic component.  
\end{itemize} 

In addition, we use theoretical model templates of \citet{Kovacevic2010} to reproduce Fe {\small II} emission when it is detected. 
Prior to the modelling of the emission lines, we
estimate the local continuum by fitting a power-law to the spectra at
both sides of the regions using those wavelength ranges that are not
affected by prominent features and or bad sky-subtraction residuals
(e.g. $4000-4050$ and $5600-5650$ for the H$\beta$ region). For each sets
of Gaussian profiles, the wavelength separation between emission line
within a given set of Gaussian is constrained in accordance with
atomic physics. This means that the velocity offset of the OC from the Narrow (systemic)
components are constrained to be the same for all the emission lines. Moreover, the relative flux of the two [N {\small II}] and [O {\small III}] components is fixed to 2.99 and [S {\small II}] flux ratio is required to be within the range $0.44<f(\lambda6716)/f(\lambda6731)<1.42$ (\citealt{Osterbrock2006}).

The different ionization potentials and critical
densities of NLR line features suggest a gas stratification with
respect to the central SMBH (\citealt{DeRobertis1986}). In order
to account for such effect, the fitting procedure would need to consider a
different line width for each emission feature within a given set of
Gaussian lines. Our fitting procedure has proved to well reproduce all
the analysed emission lines with the constraints above mentioned, from
low ionization emission lines, such as [S {\small II}] doublet (with IP $=10.4$
eV), to high ionization lines (i.e. [O {\small III}],
with IP $=35$ eV; see also \citealt{Lanzuisi2015}). Indeed, we note that even when different line widths
have been derived from low-to-median wavelength resolution data, their
values were always consistent within the errors of the Gaussian fit
(see \citealt{DeRobertis1986,WangXu2015}). We therefore
conclude that the spectral resolution of the spectra we analysed do not permit to unveil a possible NLR stratification.

\begin{figure*}[t]
\centering
\includegraphics[width=20.8cm]{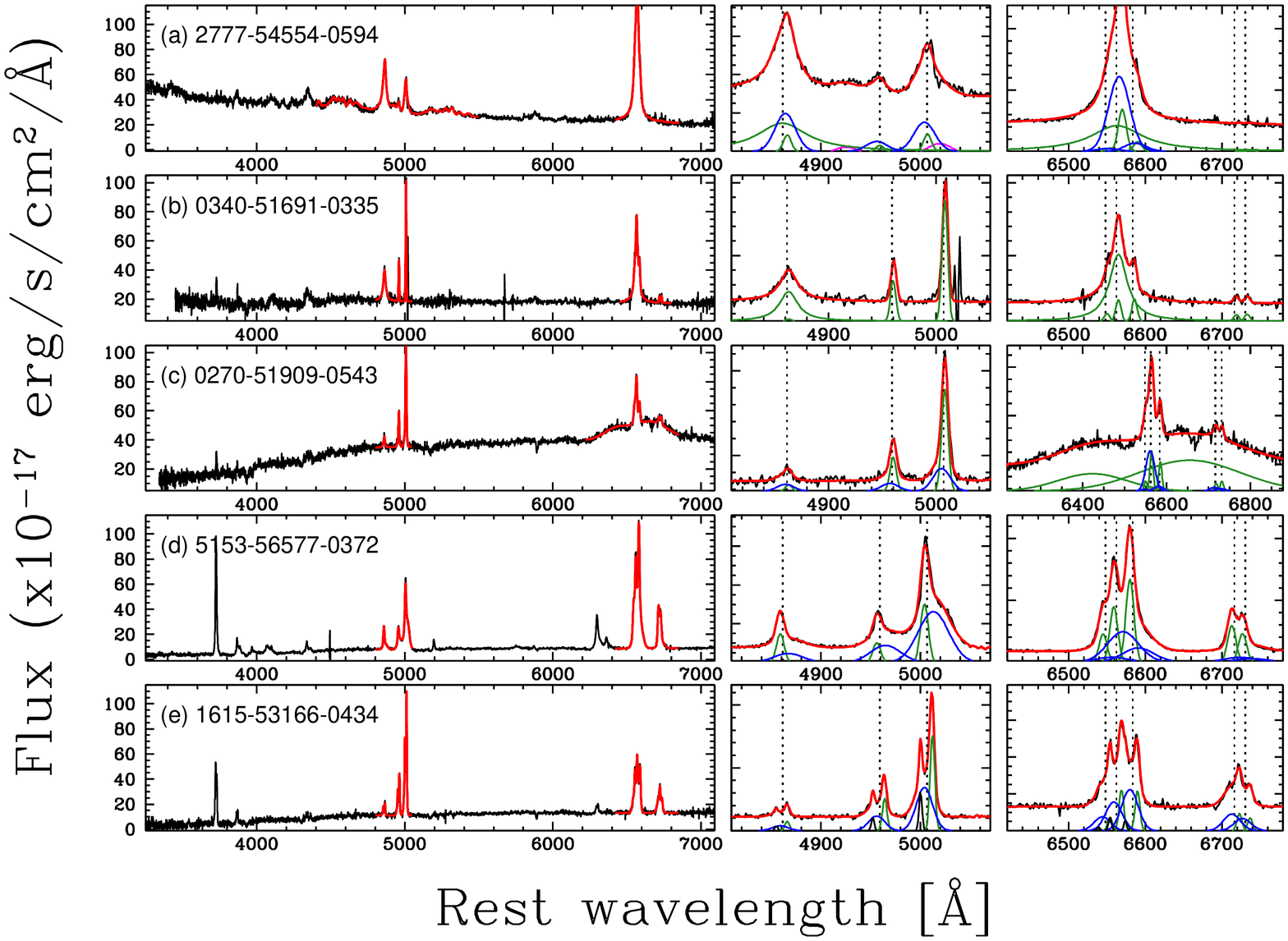}
\caption{\small Five examples of spectra illustrating our fitting method. For each object, we show on the left the spectrum (black curve) and the best-fit model (red curve). Central and right panels show the zoomed regions around the H$\beta$ and the H$\alpha$ emission, respectively. The dashed vertical lines mark the location of H$\beta$, [O {\small III}] doublet, H$\alpha$, [N {\small II}] and [S {\small II}] doublets. Best-fit NC and BC profiles are highlighted with green curves (panels $a$ to $e$); OC and Fe {\small II} emission are shown with blue (panels $a$ to $e$) and magenta (panel $a$) curves respectively. Finally,  black Gaussian profiles (in panel $e$) show a second set of NC used to fit doublet peaked galaxies.
In the rows are shown the spectra of various AGN types. From top to bottom: (a) a blue spectrum of a type 1 AGN, with broad BLR Lorentzian profiles, Fe {\small II} emission and blue wings associated with outflows; (b) a reddened type 1 AGN without evidence of outflow; (c) a low-luminosity Broad Line AGN, which continuum is dominated by stellar emission. For this target, we observe a complex BLR H$\alpha$ profile, modelled with two strong BC, and no BLR H$\beta$ emission. (d) a candidate type 2 AGN, without evidence of BLR emission but with strong emission from outflowing gas detected for all the optical emission lines; (e) a candidate type 2 AGN, with double peaked emission lines and an outflow component.}
\label{fitspettri}
\end{figure*}


\subsection{Non-parametric measurements}

We use non-parametric velocity estimator analysis \citep[see, e.g.,][]{Zakamska2014,Liu2013,Rupke2013} for the [O {\small
  III}] emission line only, to derive the kinematic conditions of the ionized gas
within the AGN NLRs. 

Non-parametric measurements are obtained by measuring velocity $v$ at which a given fraction of the line
flux is collected, using the cumulative flux function
$F(v)=\int_{-\infty}^{v} F_v(v')\, dv'$. The position of $v=0$ of the
cumulative flux is determined using the systemic redshifts derived
from the NC of the simultaneous best-fit results.
Following the prescription indicated by \cite{Zakamska2014}, we
estimate for the total emission line best-fit profiles the following parameters:
\begin{enumerate}[label=(\roman*)]
\item 
The line width $W80$, the width comprising 80\% of the flux, that for a Gaussian profile is very close to the FWHM value. It is defined as the difference between the velocity at 90\% ($v90$) and 10\% ($v10$) of the cumulative flux, respectively;\\
\item 
 The maximum velocity parameter $V_{max}$, defined as $v02$ when blue prominent broad wings are present, or as $v98$ when red ones are, on the contrary, present.
\end{enumerate}
In contrast to $V_{max}$, values of $W80$ include only differences between velocities and do not depend on
the accurate determination of the systemic velocity. A possible
residual error in the determination of the systemic velocity, however,
may produce a variation of at most few tens of km/s in the
$V_{max}$ value, corresponding to variation of a few \% for velocities higher than 500 km/s (see, e.g., Fig. \ref{LvsV}). 

The best-fit line profile used to derive non-parametric estimates is taken from the simultaneous fitting procedure results. Such approach is, in fact, less dependent on particular poor S/N conditions in the vicinity of the emission line: the low S/N regions around an individual emission feature, if fitted independently, may be wrongly interpreted as extended wings and modelled with a faint and broad Gaussian profile. 

We  fit the Gaussian profiles using a fortran code implementing the
Minuit package (\citealt{James1975}). As a first step, we fit all emission lines with minimum number of gaussian components (ie., NC, and NC+BC for broad line AGNs). The visual inspection of fit results allows us to assess the necessity of a second fit with additional OC components when significant residuals are present. Then, we accept the NC+OC(+BC) fit decomposition if it resulted in a significant improvement in $\chi^2$, with $\Delta \chi^2 \gtrsim 1000$ (reduced $\Delta \chi^2 \gtrsim 1$), i.e. $\gtrsim 3\sigma$. Our quality-of-fit criterion is justified by the large number of degrees of freedom ($\sim 1000$) of simultaneous multicomponent fit. In fact, the number of parameters we used in BC+NC or BC+NC+OC does not determine significant variations in $\chi^2$ distribution, and standard confidence intervals (\citealt{Press1992}) can be used to asses the possible improvement due to the addiction of OC Gaussians.

\begin{figure*}[!t]
\centering
\includegraphics[width=17cm,trim=65 0 0 0,clip,angle=180]{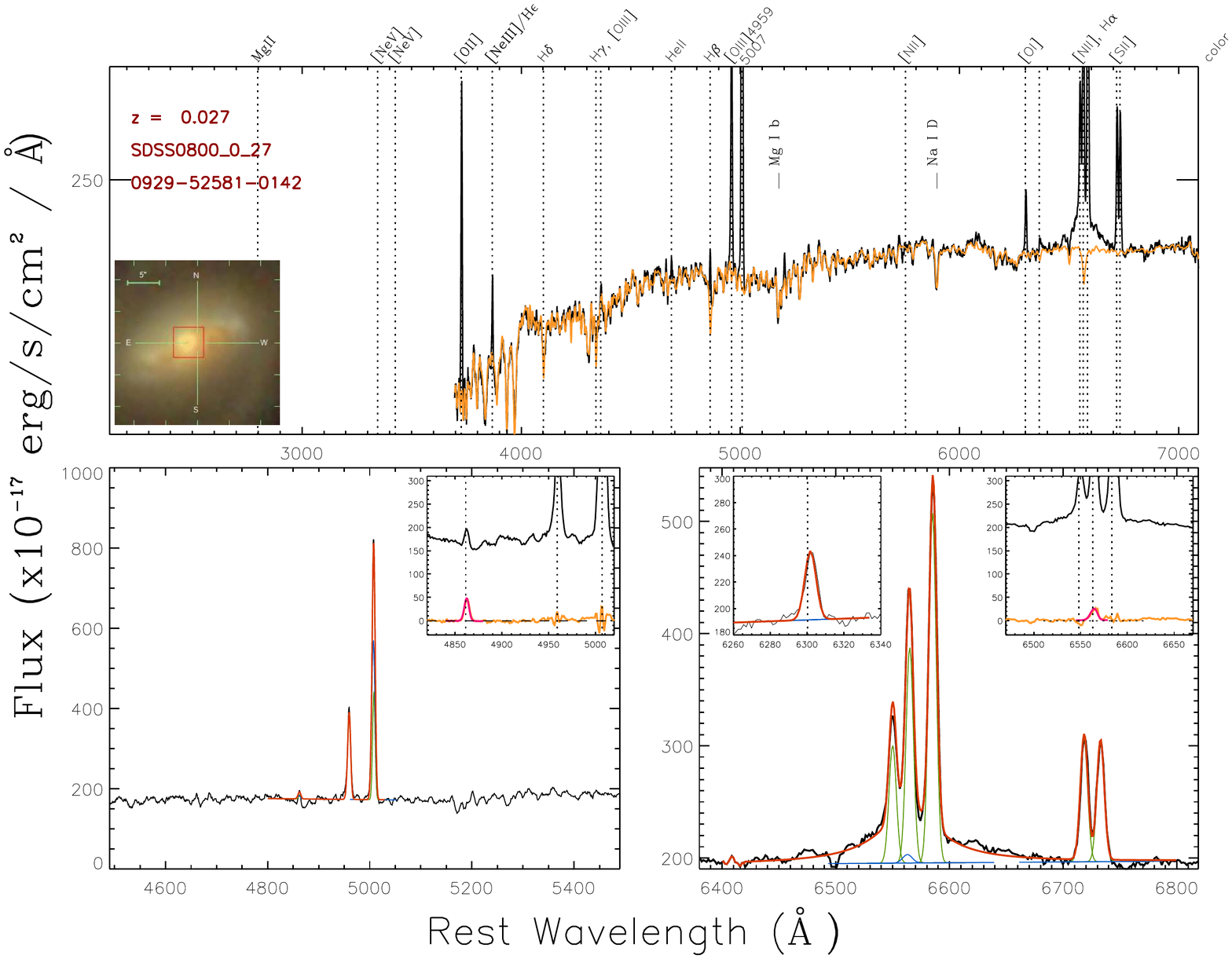}
\caption{\small ({\it top panel:}) pPXF best-fit model results for a type 1.9 AGN (orange curve) super-impressed on the rest-frame SDSS spectrum. All the prominent emission lines and the analysed features are highlighted with vertical dotted lines; the positions of Mg {\small I} b and Na {\small I} D absorption lines are also marked. The inset on the left shows the SDSS colour-composite cut-out of the galaxy, with a red square marking the spatial region from which the $3''$ spectrum has been obtained. 
({\it bottom panels:}) Fit results from the multicomponent simultaneous fit in the H$\beta$-[O {\small III}] (left) and H$\alpha$-[NII] (right) regions. Green and blue curves show the NC and OC Gaussian profiles, while the red lines represent the total best-fit profile. The insets on the right of each panel show the excess in the Balmer emission lines found after the correction for the pPXF best-fit model (orange curves, representing the difference between the SDSS spectrum and pPXF {\it and} simultaneous fit models). The original spectrum is also shown for a qualitative evaluation of the contribution of stellar absorption feature in the observed emission line. Magenta Gaussian profiles represent best-fit results associated with significant ($>$5$\sigma$) residuals. The insets on the left of the H$\alpha$+[N {\small II}] panels show the best-fit models of [O {\small I}]$\lambda$6300. 
}
\label{ppxf}
\end{figure*}

In order to estimate errors
associated with our measurements, we use Monte Carlo simulations. For
each modelled spectrum, we collect the fit results of 30 mock spectra obtained from the best-fit
final models (red curves in Fig. \ref{fitspettri}, \ref{ppxf}), and adding Gaussian
random noise (based on the standard deviation of the corresponding
local continuum). 
When Fe {\small II} features were fitted, prior to the Monte Carlo simulations, we subtracted our best-fit Fe {\small II} template in order to minimise the degeneracy in the errors estimation. 
The errors are calculated by taking the range that contains 68.3\% of values evaluated from the obtained
distributions for each Gaussian parameter/non-parametric measurement.
Finally, we note that since line profiles generally are non-Gaussian and much broader than the spectral resolution (see below), we do not correct the observed profiles for instrumental effects and report all values as measured.

\begin{figure*}[t]
\centering
\subfloat[]{\includegraphics[width=8cm,trim=0 130 0 140,clip]{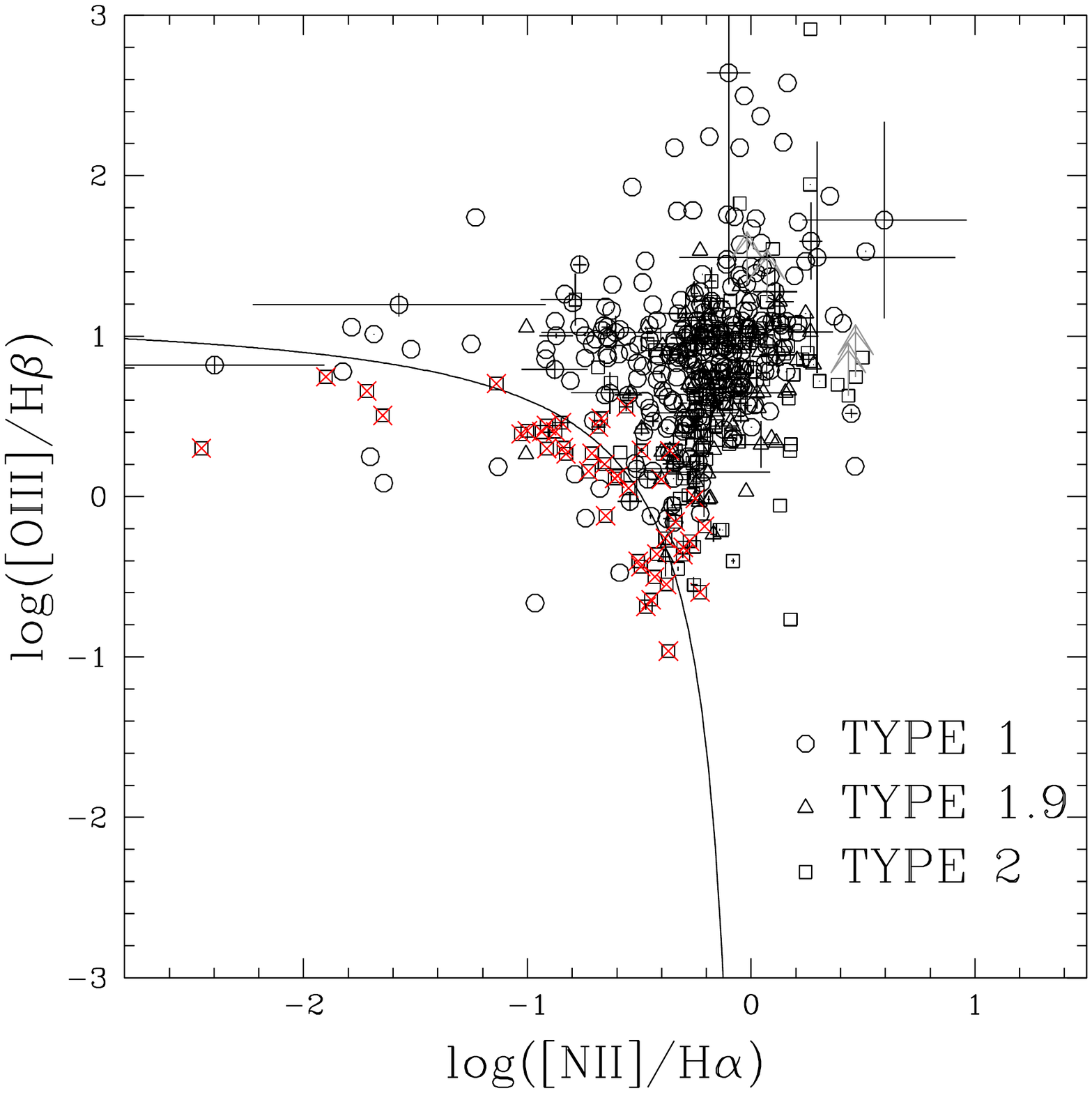}}
\hspace{-1cm}
\subfloat[]{\includegraphics[width=8cm,trim=0 130 0 140,clip]{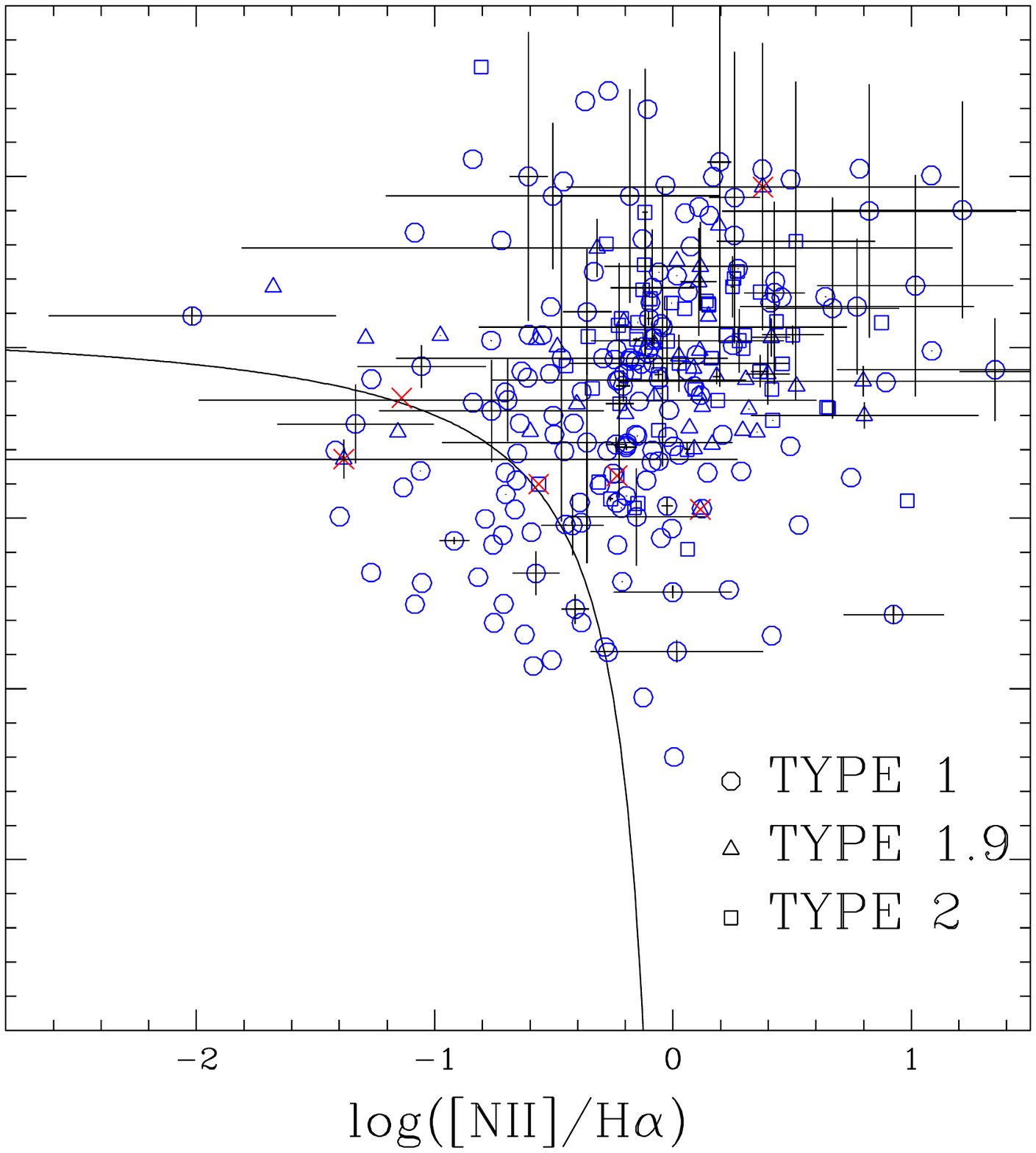}}
\caption{\small BPT diagrams -- Standard  diagnostic  diagram  showing  the  classification scheme by \citet{Kewley2013}. The lines drawn in the diagram  correspond  to  the  theoretical  redshift-dependent  curves at z $=0$ used  to  separate  purely  SF  galaxies  from  galaxies  containing AGN. Black (left panel) and blue (right panel) symbols correspond to the systemic NC and outflow OC flux ratios, respectively. Representative error bars are shown only for a small fraction of targets. Circles, squares and triangles denote Type 1, type 2 and type 1.9 AGNs respectively. Red crosses highlight the SF galaxies discarded from the sample, using the NC flux ratios (see text for details). Grey upper arrows represent lower limits, due to undetected H$\beta$ emission line.}
\label{bpt}
\end{figure*}

\subsection{Spectral fit results}

Figure \ref{fitspettri} plots five examples of different types of
sources found in our sample: blue ({\it a}) and red
({\it b})  spectra of type 1 AGNs, in which Fe {\small II} emission and/or BC could be present; low-luminosity type 1 AGNs which continuum is dominated by stellar emission ({\it c}); red spectra in which the same asymmetric profile is found in all optical emission lines and is therefore associated uniquely to OC, without clear sign of BLR Balmer emission ({\it d}); objects with double peaked emission lines, modelled with two NC together with an OC ({\it e}).

We note that these AGN spectra often present significant Fe {\small II} emission. As described in Appendix \ref{AppendixA}, such emission is strongly blended with BLR H$\beta$ emission and the red wing of the [O {\small III}] line. As a result, in order to derive correct BLR profiles and non-parametric kinematics for the doubly ionized oxygen, an adequate modelling of the iron emission is required. We use theoretical model templates of \citet{Kovacevic2010} to reproduce such emission. Two representative best-fit results showing strong blending with both [O {\small III}] and H$\beta$ lines are shown in Fig. \ref{feiisubtraction}.

Through our own line fitting routine we found 441 broad line (BL) AGNs, comprising both type 1 and type 1.9 sources (showing BC in the H$\alpha$ and not in higher order Balmer lines; see Fig. \ref{fitspettri}, panel $c$), and 165 type 2 AGN candidates.
We excluded from any further analysis 7 [O {\small III}] double peaked galaxies\footnote{Double peaked profiles could be associate both with bi-conical QSO winds and binary AGNs; SDSS spectra do not allow a separation between the two classes of objects (see discussion in \cite{Yuan2016}, sec. 3.2) and are therefore excluded.} (see, e.g.,  Fig. \ref{fitspettri}, panel $e$; Fig. \ref{feiisubtraction}, top panel), 5 red galaxies with high sky residuals at $\approx 5000\AA$ responsible for bad estimate of the S/N in the {[O {\small III}] region, and 6 galaxies with strong and complex stellar continuum in the proximity of H$\alpha$ region.

\subsection{BPT emission line diagnostics}  

The presence of BLR emission represents an unambiguous indicator for AGN presence in the nucleus; however, some of our targets show concomitant stellar continuum and BLR features, indicating the presence of low-luminosity (or obscured) AGNs. 
Precisely, about 20\% of our BL AGNs are type 1.9 sources.
In order to confirm the nature of the AGN candidates and to discriminate between SF and AGN photo-ionized emission lines in faint type 1.9 and type 2 AGNs, we use the optical BPT diagnostic diagram (\citealt{Baldwin1981}) as a further tool to investigate the nature of the ionizing sources. 

Prior to compute the line flux ratios, we fit the stellar continuum using penalised pixel fitting (pPXF; \citealt{Cappellari2004,Cappellari2016}) and correct the Balmer line fluxes taking into account the stellar features from the pPXF best-fit model. In fact, underlying stellar absorption of the Balmer lines are expected to be not negligible in low-luminosity AGNs and to shape the emission line profiles. pPXF routine is a program developed by Cappellari et al. to extract the galaxy stellar kinematics (i.e., stellar velocity dispersion $\sigma_*$) from absorption-line spectra;
to adopt the procedure for BL AGNs, a window of $1.2\times10^4$ km/s around the expected position of permitted emission lines is excluded from the fit. pPXF method is able to well reproduce the continuum emission of $\sim$ 35\% of the sample (we discarded all best-fit results associated with errors $>$20\% in velocity dispersion parameter).
Figure \ref{ppxf} shows an example of a pPXF best-fit model for a type 1.9 AGN candidate (orange curve in the top panel). 
In the bottom panels we show the results obtained from the multicomponent simultaneous fit (red curves) and, in the insets, the excess in the Balmer emission lines found after the correction for the pPXF best-fit model. We found that the Balmer stellar absorption features can actually determine underestimates in the fluxes, with median values of $\approx$ 5\% and $\approx$ 20\% of the H$\alpha$ and H$\beta$ respectively. We note that, as expected, such absorption features affect only the narrow emission components.

Figure \ref{bpt} shows the BPT diagrams obtained from our spectroscopic analysis, after the correction from stellar features, for both NC (left) and OC (right). The lines drawn in the diagrams correspond to the theoretical curve (at z $= 0$) used  to  separate  purely  SF  galaxies  from  galaxies  containing AGN (Eq. 1 of \citealt{Kewley2013}).

The use of BPT diagrams for both NC and OC is justified by recent results obtained with spatially resolved spectroscopy, which probe the different locations and ionisation conditions of unperturbed and outflowing gas  (e.g., \citealt{Arribas2014, Cresci2015,Cresci2015b,Kakkad2016, McElroy2015,  Perna2015a,Perna2015b, Westmoquette2012}).

For almost all the sources, the systemic NC and the OC are consistent with an AGN classification. OC measurements are more scattered because of their associated lower intensities: this determine an important degeneracy in the fit results, in particular for type 1 AGNs (blue circles) for which a higher number of components must be taken into account (namely: NC, OC and BCs). 

Because of the mentioned arguments, we excluded from the following analysis 43 targets (marked with red crosses in the figure) using only NC flux ratios. 
For these sources, the SF nature highlighted by the BPT diagrams, has been confirmed by the concomitance of red spectra and low X-ray luminosities (i.e. $<10^{42}$ erg/s in the $2-10$ keV band). We note that the exclusion of few targets above the theoretical transitional curve is due to a conservative approach that takes into account:
\begin{itemize}
\item a possible stellar absorption feature contribution in those sources for which low S/N spectra do not allow stellar features modelling, but for which we expect some contribution (i.e., type 1.9 and type 2 sources). 
In order to take into account this effect, we derive representative shifts in the BPT diagram due to possible underestimates in Balmer flux estimate, assuming that all H$\alpha$ and H$\beta$ fluxes are increased by a factor of 3$\times$ the median  pPXF corrections mentioned above.
These 'corrections' correspond to a downward shift of $\approx 0.2$ in log([O {\small III}]/H$\beta$), and a left shift of $\approx$ 0.1 in log([N {\small II}]/H$\alpha$).
\item the error bars associated with each source, due to strong degeneracy in the fitting procedures when low S/N spectra are analysed. 
\end{itemize}

About 14\%  of our targets have z $>0.4$ and their H$\alpha$ region is not covered by the SDSS spectra. For these targets it is therefore not possible to use the BPT diagnostic. 
However, they are generally associated with blue spectra, and 95\% of them show unambiguous BLR H$\beta$ emission. 
For the remaining 5\% (4 targets) we observed [O {\small III}]/H$\beta$ consistent with the average value observed for the entire sample (log[O {\small III}]/H$\beta\sim1$), for both NC and, when present, OC. We therefore confirm the AGN nature for all the z $>0.4$ targets. 

Summarizing, thanks to the BPT diagnostic coupled with a visual inspection of the spectra and the available X-ray analysis, we obtain a final sample of 563 AGNs (362 type 1, 77 type 1.9, 124 type 2).

\begin{figure}[t]
\centering
\includegraphics[width=8.8cm,trim=0 130 20 90,clip]{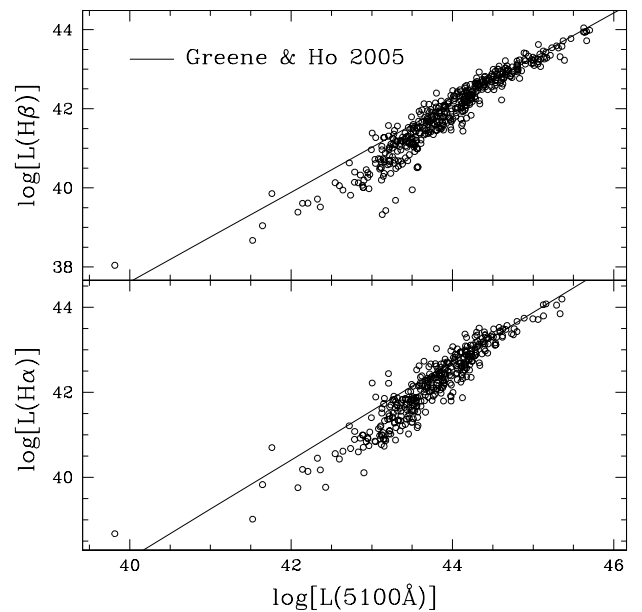}
\caption{\small $L_{H\alpha}-$ (bottom panel) and $L(H\beta)-L(5100\AA$ (bottom) correlations. The solid lines represent the \citet{Greene2005} correlations obtained from a sample of type 1 AGN with low galaxy contribution in the optical continuum emission. A significant displacement from such relations is found for our low-luminosity AGN sub-sample, in which the continuum emission at $5100\AA$ is dominated by host galaxy emission.
}
\label{5100balmer}
\end{figure}


\begin{figure}[b]
\centering
\includegraphics[width=9cm,trim=0 130 20 90,clip]{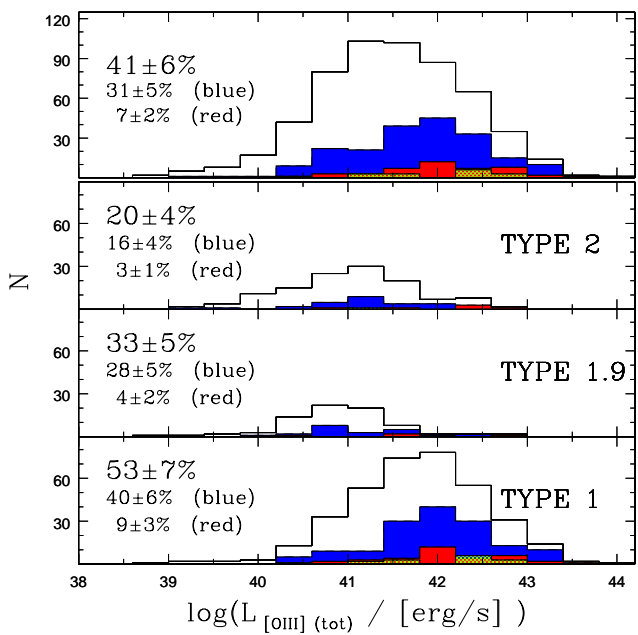}
\caption{\small Distributions of sources with ionized outflows as a function of the [O {\small III}] luminosity. Top panel show velocity distribution for approaching, receding and symmetric outflows, represented with blue, red and green histograms. The lower panels show same distributions for sub-samples defined on the basis of the AGN type. For each panel, the fraction of AGNs with outflow is highlighted, together with the fraction of receding and approaching outflows.}
\label{Outflowdistribution}
\end{figure}

\section{The nuclear properties}\label{nuclearSDSS}
We use single-epoch (SE) technique to determine the BH masses of our sub-sample of BL AGNs. Such technique combines the virial theorem with the BLR radius - luminosity relation (see, e.g., \citealt{Kaspi2005,Bentz2013,Saturni2016} ), allowing an estimate of the black hole mass from measurements of the BLR line width and the AGN luminosity. Different empirical relations have been calibrated in the last years (e.g., \citealt{Vestergaard2006,Shen2011,Bongiorno2014}) for various BLR line width and AGN luminosity indicators. Here we adopt the SE relations calibrated by \citet{Greene2005},
\begin{equation*}
log \left ( \frac{M_{BH,SE}}{M_{\odot}} \right)= 6.30_{-0.07}^{+0.08}+
\end{equation*}
\begin{equation*}
(0.55\pm0.02)\  log \left ( \frac {L(H\alpha)}{10^{42} erg/s} \right ) +(2.06\pm0.06)\ log\left(\frac{FWHM} {10^3 km/s}\right),
\end{equation*}
\begin{equation*}
log \left ( \frac{M_{BH,SE}}{M_{\odot}} \right)= 6.56_{-0.03}^{+0.02}+
\end{equation*}
\begin{equation}\label{SEgreene}
(0.56\pm 0.02)\  log \left ( \frac {L(H\beta)}{10^{42} erg/s} \right ) +2\ log\left(\frac{FWHM} {10^3 km/s}\right).
\end{equation}
These relations replace the $\lambda L_\lambda$ at $5100\AA$ ($L(5100\AA)$ hereinafter) usually associated to SE relations calibrated on Balmer lines (\citealt{Shen2011}), with their BLR luminosities, and are based on the tight correlations 
found between Balmer and continuum luminosities in type 1 AGNs with strong blue thermal nuclear continuum (see \citealt{Greene2005}, fig. 2). We adopt such relations in order to mitigate the effects of host galaxy contamination at $5100\AA$ in our sample. 
Figure \ref{5100balmer} shows the Balmer luminosities against the $5100\AA$ continuum luminosity for our BL AGNs. The two panels display a significant displacement from the Greene and Ho relations at lower Balmer luminosities ($L(H\alpha)$ and $L(H\beta)\lesssim 10^{42}$ erg/s), due to the enhancement in the continuum luminosity $L(5100\AA)$, which is strongly contaminated by the host galaxy emission.

In order to derive the black hole masses, we adopt the FWHM measured from the best-fit model of the BLR profile as line width (see, e.g., \citealt{Shen2011}). Such procedure takes into account the additional multiple Gaussian (and/or Lorentzian) components required to reproduce peculiar broad H$\alpha$ and H$\beta$ BLR line profiles (see, e.g., Fig. \ref{fitspettri}, panel $c$). 
With these assumptions on AGN luminosity and FWHM, we derive black hole masses in the typical range $10^6-10^9$ M$_\odot$. 

We also estimate AGN bolometric luminosities. We note that, in order to compute an X-ray bolometric correction and test in a statistical way the results we obtained in our previous works (e.g., \citealt{Brusa2016}), we need a bolometric luminosity estimator independent from X-ray emission. In fact, contrary to the small samples at higher redshifts carefully selected in the COSMOS field and for which we could derive all the critical nuclear AGN properties through spectral energy distribution (SED) fitting procedures, this SDSS sample do not allow such accurate approach. Hence, we should refer to the SDSS spectral information. 
The unperturbed NC of the [O {\small III}] line is usually referred as a good tracer of the AGN luminosity (e.g., \citealt{Panessa2006,Jin2012}; while the OC emission may be due to different processes, such as shocks).  However, we exclude the use of [O {\small III}] line in order to reduce possible spurious correlations in our analysis (e.g., the outflow velocity estimators depend on the NC flux of the [O {\small III}] line). 
We choose to use instead the $5100\AA$ indicator, taking advantage from the above mentioned BLR-continuum luminosity relations to 'correct' the observed $L(5100\AA)$ in the low-luminosity regime, and applying the luminosity-dependent bolometric correction$\footnote{Bolometric corrections convert a luminosity at a specific wavelength to $L_{bol}$, considering typical AGN SED.}\  $b(5100\AA) = 53-log[L(5100\AA)], presented by \citet{Netzer2014}\footnote{ Unperturbed NC doubly ionized oxygen and host-galaxy-contribution corrected continuum luminosities are well correlated (Spearman ratio of 0.85, with null hypothesis probability $< 10^{-5}$). }.

\section{Incidence of ionized outflows}\label{sdssincidence}

\begin{figure*}[th]
\centering
\includegraphics[width=19cm,trim=0 0 0 0,clip]{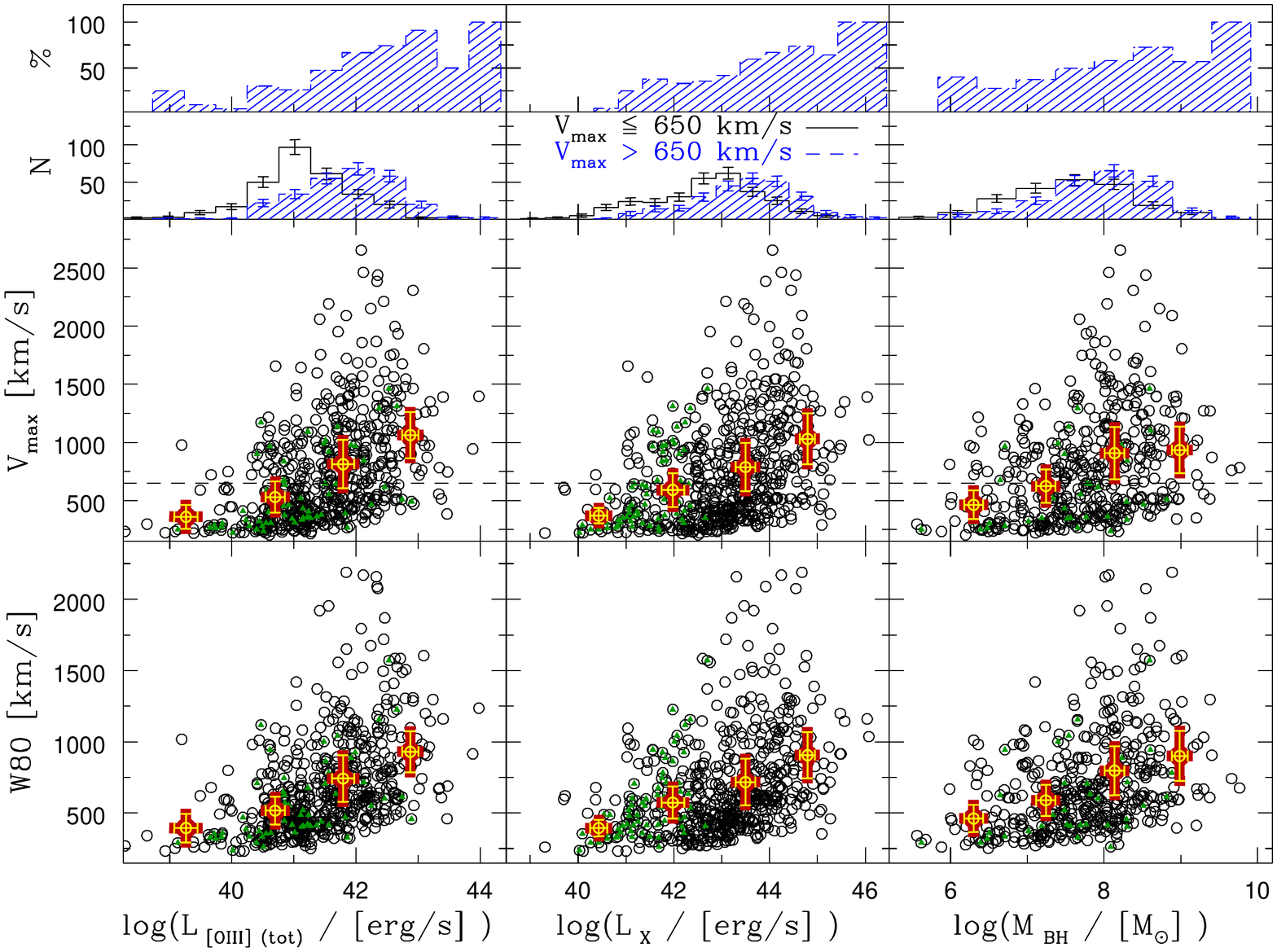}
\caption{\small ({\it central and bottom panels:}) Non-parametric velocity $W80$ and $V_{max}$ estimators as a function of [O {\small III}] (left) and intrinsic 2-10 keV X-ray (central) luminosity, as well as a function of the black hole mass for the BL AGN sub-sample.  Red/yellow dots represent average velocity in bin of luminosity; bars indicate uncertainties of the averages. Green dots mark faint X-ray sources for which the intrinsic $L_X$ has been inferred using HR ratio corrections (see text). ({\it top panels:}) Fraction of AGNs with $V_{max}>650$ km/s (blue dashed line; first panel from the top) and distributions of $V_{max}<650$ km/s (black line) and $V_{max}>650$ km/s (blue dashed line; second panel from the top) as a function of [O {\small III}] (left), X-ray (central) luminosity and $M_{BH}$ (right). }
\label{LvsV}
\end{figure*}

In order to  discriminate between gravitational and outflow processes, we choose a conservative velocity threshold $V_{max}=650$ km/s. This value corresponds to the maximum velocity derived for a Gaussian profile with a velocity dispersion of 340 km/s, and has been chosen considering that 95\% of massive BOSS galaxies below z $=0.8$ have velocity dispersions lower than this value (\citealt{Thomas2013}).

Critical broadening in forbidden emission lines can be originated also by merging events or inflowing gas. We stress however that maximum velocities higher than  $\sim$ 600 km/s are not common even in sub-millimeter galaxies at z $\sim 2$, were we expect to observe ongoing mergers (see, e.g., \citealt{Collet2016}), or in inflowing gas (e.g., \citealt{Bouche2016} and references therein).

 We assume that $V_{max}$ is representative of the outflow velocity (assumed to be constant with radius). From a geometrical point of view, we consider $V_{max}$ as proper of emitting gas close to our line of sight, and that all lower velocities observed in oxygen line profile are due to projection effects (see \citealt{Cresci2015}). 
These assumptions, generally adopted in literature (e.g., \citealt{Brusa2016,CanoDiaz2012,Kakkad2016}), are required because of the still unsettled geometry of the outflow (\citealt{Carniani2015,Cicone2014}).

Figure \ref{Outflowdistribution} (top panel) shows the distribution of sources against [O {\small III}] luminosity for the full sample and for three sub-samples defined on the basis of the presence/absence of blue or red prominent wings in the [O {\small III}] profile: we found signature of outflows in $\approx 41\%$ of AGNs, of which 31\% and 7\% have respectively blue and red prominent wings, while only  3\% show symmetric profiles. Lower panels in the figure show the distributions for given spectral type, i.e. splitting the sample in type 2, type 1.9 and type 1 AGNs. We note that the fraction of outflows increases going from type 2 to 1, up to over $50\%$ in type 1 AGNs. In particular, we note that the relative fraction of blue/red outflows are roughly similar in each spectral type: the fraction of incoming outflows are always $\sim$ 5 times that of receding outflows. In the context of the unified model, in type 1 AGNs the torus structure should force the outflowing gas toward our direction, i.e. we should be able to observe more sources with blue outflows. On the contrary, in type 2 AGNs, the torus axis should be perpendicular to our line of sight and the ejected material should emerge along that direction (the projected outflow velocities in this case are smaller, easily explaining the lower number of sources with outflows in type 2 AGNs). 
Naivily, one would expect to observe a larger fraction of symmetric outflows in type 2 AGNs. However, the dust distribution in the disk and the inclination of the host galaxy respect to the line of sight  is probably enough to obscure a large fraction of the flux from the receding outflow, regardless the AGN spectral classification  (see, e.g., \citealt{Fischer2013,Muller2011,Bae2014}).

Figure \ref{LvsV} shows the distributions of the two velocity estimators against the total [O {\small III}] (on the left) and the intrinsic X-ray $2-10$ keV (centre) luminosities, as well as the black hole masses for the BL AGN sub-sample (right panels). We discuss here the major results for each pair of indicators: 

\vspace{3pt}
 $V_{out}$ vs. $L_{[O {\small III}]}$:  Both non-parametric estimators show a positive correlation with increasing [O {\small III}] luminosity: moving from lower to higher luminosities, an increasing number of targets exhibit higher velocities. A clear trend appears when we consider the average velocities in bin of luminosity (red-yellow points in the lower and central panels). The same positive trend appears also when we consider median instead of average values - or when we combine our results with those available from literature (see Appendix \ref{AppendixC}).
We also investigate how the fraction of AGNs with/without outflows changes as a function of the $L_{[O {\small III}]}$ (per luminosity bins; first row in the left panel). 
We found that above $L_{[O {\small III}]}\approx10^{42}$ erg/s the fraction of AGNs with outflows becomes $> 50\%$, as also highlighted in the second row first (left panel), where we show  the number of AGNs with V$_{max}$ above/below the velocity threshold value. Assuming a bolometric correction of $\sim3\times10^3$ for the [O {\small III}] line (\citealt{Heckman2004}), this corresponds to a bolometric luminosity of $\approx10^{45}$ erg/s, which is consistent with the luminosity threshold proposed by \citet{Veilleux2013}, obtained studying the incidence of molecular outflows in ULIRGs hosting AGNs, and by \citet{Zakamska2014} and \citet{Woo2016}, derived analysing ionized outflows in obscured and type 2 QSOs.

\vspace{3pt}
 $V_{out}$ vs. $L_X$: The same results are found when we consider the X-ray luminosities (central panels): a clear trend with increasing $L_X$ is observed, both with $W80$ and $V_{max}$ measures. Also in this case, the fraction of sources with outflows is higher in the high-luminosity regime (upper panel). 
 We underline that this is the first time we observe such correlations between X-ray activity and outflow effects in a large sample, confirming the results obtained on a smaller sample of z $\sim 0.6-1.6$ by \cite{Harrison2016} and based  only on 2 luminosity bins (see their fig. 8).
We note that in the low-luminosity regime a small group of sources are displayed with green symbols. 
For these sources we cannot derive the intrinsic X-ray luminosity from proper X-ray spectral analysis because associated with poor detections (i.e., $<100$ counts in the $2-10$ keV band). For these sources we show therefore the X-ray luminosities as inferred from the hardness ratio (HR) correction (following \citealt{Merloni2014}). 
Being a large fraction of these sources associated with high HR, we suggest that, if properly corrected for extinction, these sources could even reinforce the correlation between X-ray luminosity and outflow velocity.

\vspace{3pt}
 $V_{out}$ vs. M$_{BH}$:  We found the same positive correlations also in the right panels. This trends are expected since the M$_{BH}$ is derived from Eq. \ref{SEgreene}, i.e. is proportional to a third indicator of the bolometric luminosity, the BLR Balmer emission. As for [O {\small III}] and X-ray luminosities, here we find a black hole mass threshold of $10^8$ M$_\odot$. 
We stress that such result allows an interesting confirmation of the evolutionary model predictions (see \citealt{King2015}): comparing the gas bulge binding energy of a typical massive galaxy of $M_{bulge}\sim10^{11}$ M$_\odot$, stellar velocity dispersion $\sigma_*\sim 300$ km/s and a gas fraction $f_{gas}=0.162$ (the cosmological mean value; \citealt{Planck2014}),
\begin{equation*}
E_{bulge}\sim f_{gas}M_{bulge} \sigma_*^2\approx 10^{59}\ erg ,
\end{equation*}
  with the predicted outflow energy computed under the assumption of a small coupling factor with the released SMBH energy ($\approx 1\%$, as we actually found in AGN-driven outflows; see \citealt{Carniani2015,Fiore2017}), and an Eddington accreting mass rate,
\begin{equation*}
E_{out}\sim0.01\times E_{BH}\approx 10^{51} (M_{BH}/M_\odot)\ erg,
\end{equation*}
we find that only when the SMBH reaches a M$_{BH}\sim10^8$ M$_\odot$ it is powerful enough to release such energy in the form of powerful outflows. Of course, we note that large uncertainties in deriving the $M_{BH}$ (see discussion in \citealt{Shen2011}), the unknown host galaxy properties and the assumed Eddington accretion onto the SMBH\footnote{From the derived bolometric luminosities and black hole masses we are able to estimate the Eddington ratio $\lambda_{Edd}=L_{bol}/L_{Edd}\propto L_{bol}/M_{BH}$ for our sample of BL AGNs. Both sources with and without outflows are characterized by average Eddington ratio of $\approx0.1$, with a possible mild evidence of higher $\lambda_{Edd}$ in those sources with signatures of outflows. We note however that any strong conclusion is avoided due to the strong uncertainties in deriving $\lambda_{Edd}$ and to AGN flickering considerations (e.g., \citealt{Schawinski2015}).} must be taken into account. For all these reasons, our result represent a mild confirmation of theoretical predictions.

A possible criticism of these $V_{out}\ -$ AGN power relations is that when $L_{[O {\small III}]}$ is small, it is more difficult to detect and model faint wings associated with outflows. However, the sources associated with smallest luminosities are also those with smallest z and highest S/N, as highlighted in Fig. \ref{zdistribution} (see also the detailed discussion in Appendix Sect. \ref{AC3}). Therefore, we can conclude that,
overall, all these correlations show an increasing incidence of outflow processes with AGN power. We use Spearman rank (SR) correlation coefficient to derive the significance of the observed trends: we find coefficients of $\approx 0.5-0.7$, with probabilities of  $\ll$ 0.001 for the correlation being observed by chance. Such correlations also explain the observed fractions of outflows found in different spectroscopic AGN types, with an higher incidence in type 1 rather than in type 2 sources because the former are characterized, on average, by higher luminosities (see Fig. \ref{Outflowdistribution}).

\begin{figure*}[t]
\centering
\includegraphics[width=9.45cm,trim=0 130 0 140,clip]{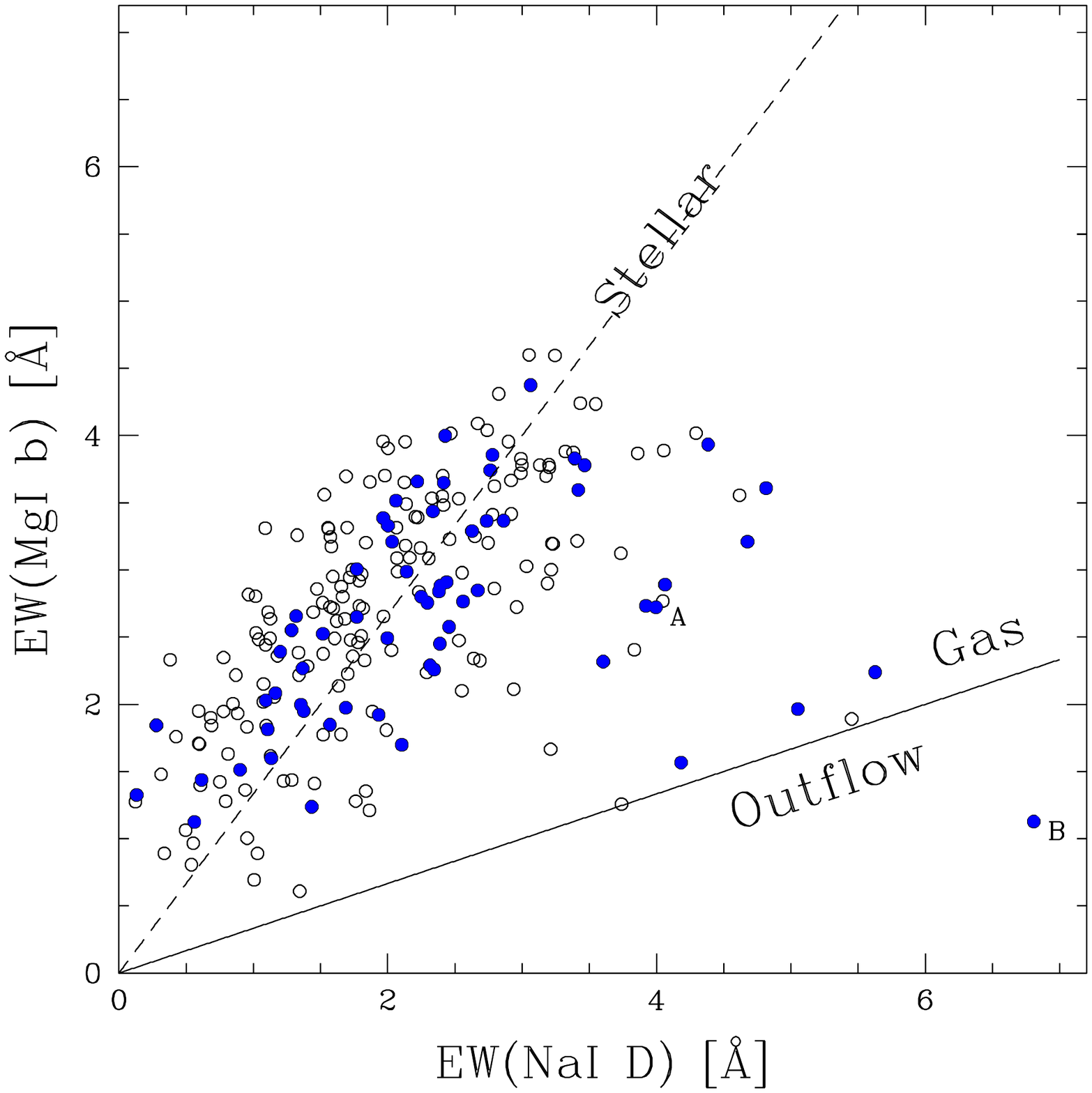}
\hspace{-0.7cm}
\includegraphics[width=9.45cm,trim=0 130 0 140,clip]{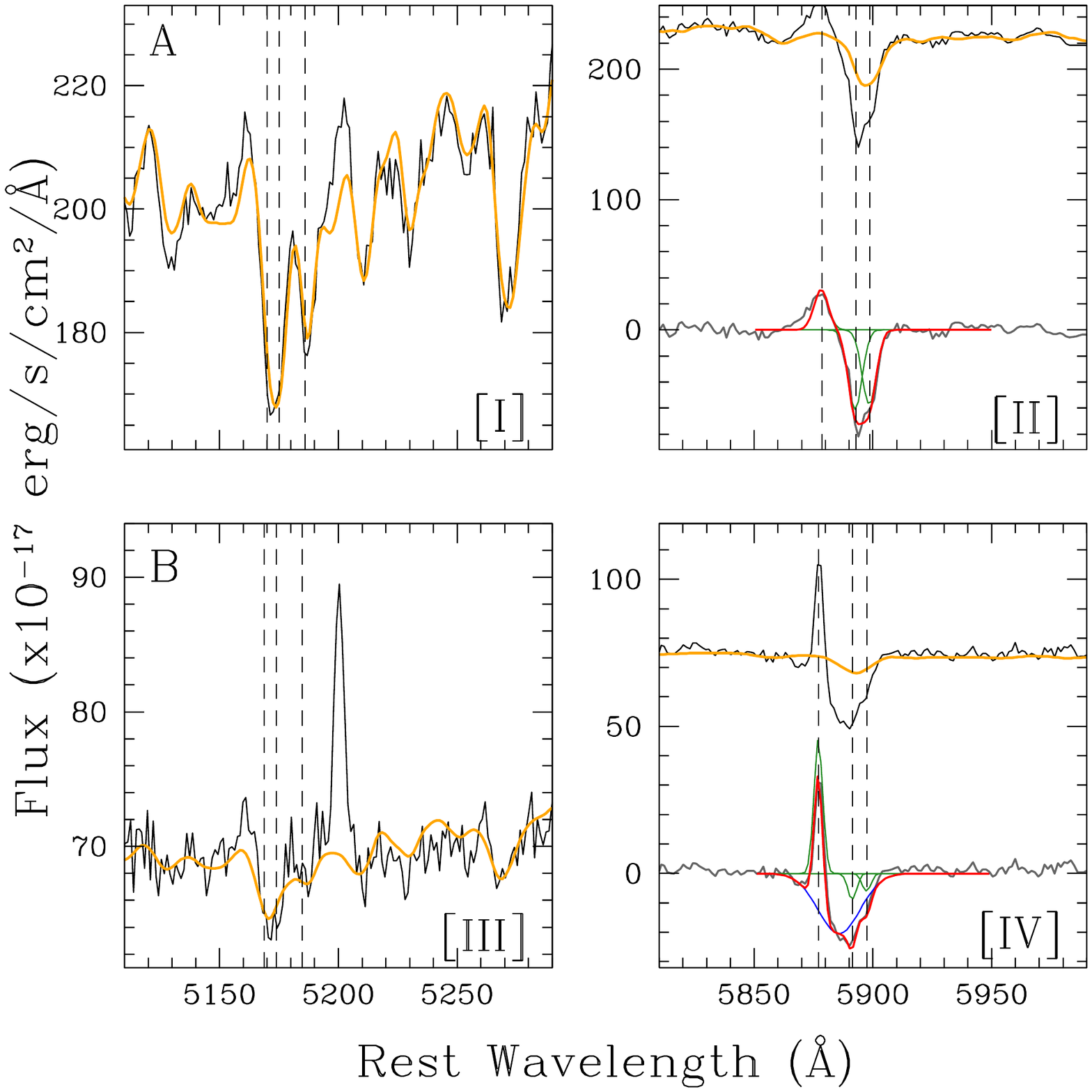}
\caption{\small ({\it  left:}) EW of the Na {\small I} D versus Mg {\small I} b absorption lines. Open black and solid blue points represents sources with and without ionized outflows respectively (i.e. V$_{max}\lessgtr$ 650 km/s). The dashed line marks the ratio at which the sodium is expected on the basis of observed magnesium EW (\citealt{Heckman2000}); the solid line marks the starburst-driven outflow region proposed by \citet{Rupke2005a}. A and B letters indicate the sources for which the spectra are shown on the right. ({\it right:}) SDSS spectra of J113240.24+525701.3 and MKN 848 (marked with A and B letters in the diagnostic diagram) around the Mg {\small I} b (left panels) and Na {\small I} D (right panels) absorption features. Orange curves show the best-fit pPXF results. Vertical lines represent the systemic of Mg {\small I} b triplet ([I] and [III] panels), and He {\small I} and Na {\small I} D doublet ([II] and [IV] panels). Panels [II] and [IV] show our fit results. Green, blue and red curves represent the NC, OC and the total best-fit model respectively.}
\label{NaIDMgIb}
\end{figure*}

\begin{figure*}[t]
\centering
\includegraphics[width=8cm,trim=0 130 0 100,clip]{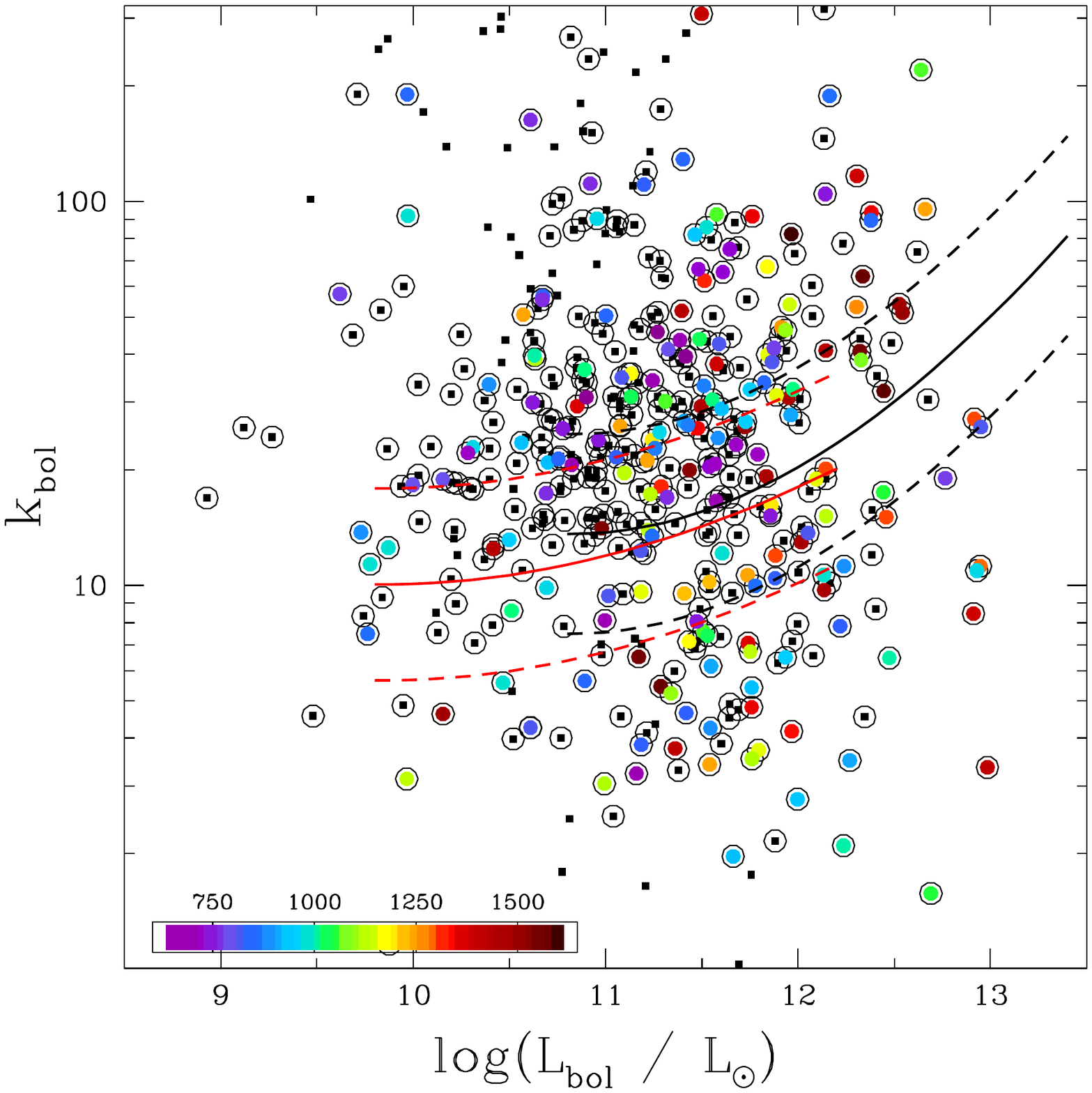}
\hspace{0.01cm}
\includegraphics[width=8cm,trim=0 130 0 100,clip]{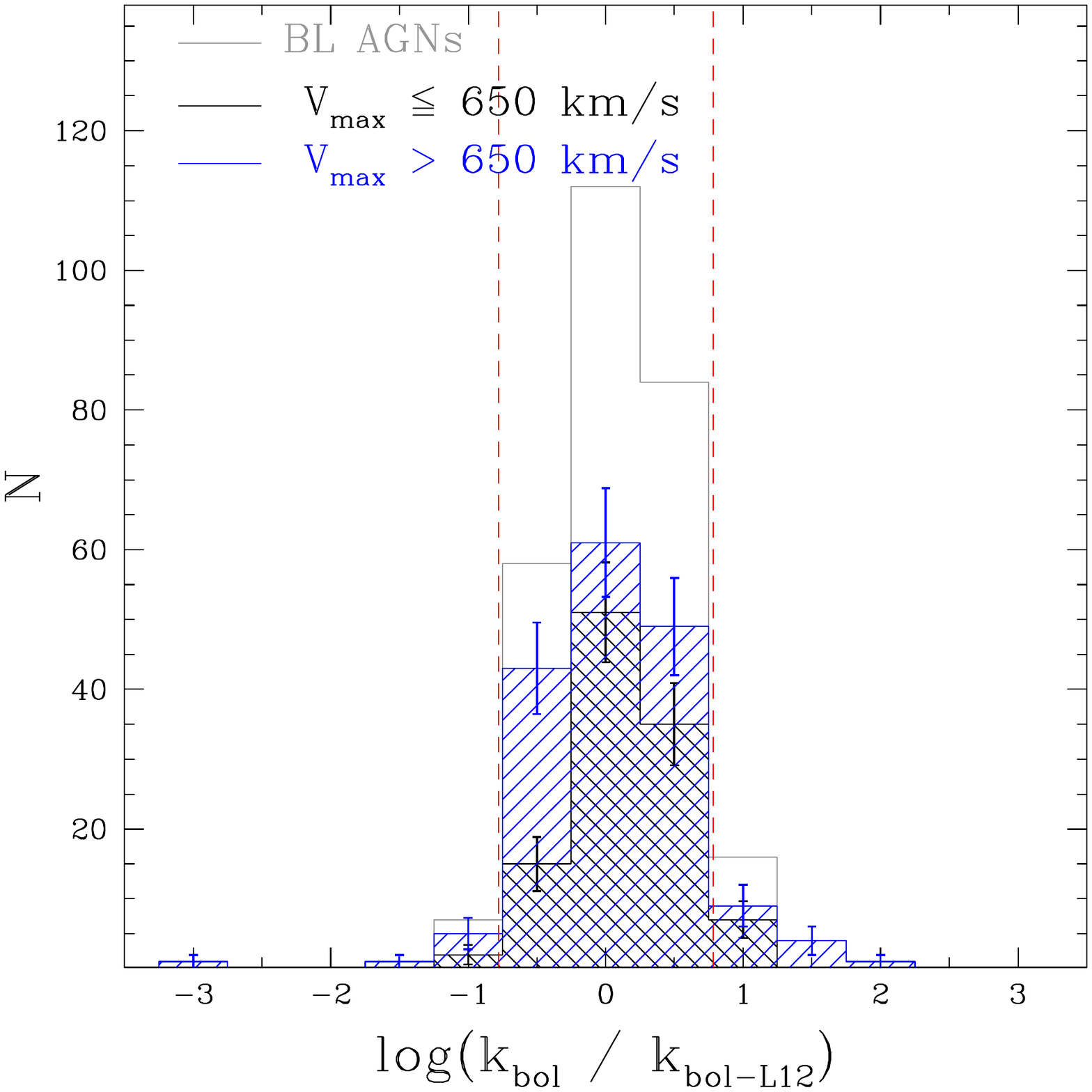}
\caption{({\it left panel:}) $k_{bol}$ vs. $L_{bol}$. Black square symbols represent type 2 AGNs, while colour-coded circles mark the type 1 and type 1.9 sources for increasing outflow velocity. The black and red curves represent the best-fitting relations obtained by \citet{Lusso2012} for type 1 and type 2 AGNs in the COSMOS field, respectively. ({\it right:}) Distribution of the normalized (in log space) $k_{bol}$ values with respect to those expected from the relation for X-ray selected AGNs (black curve in the left panel) for the BL X-ray/SDSS sample (grey histogram). Blue and black histograms represent the distributions for sources with and without outflow signatures. Error bars indicate poissonian errors. For comparison, $3\times \sigma$ dispersion values derived by \citet{Lusso2012} for XMM-COSMOS sources are also shown with red dashed vertical lines.}
\label{kbolcorr}
\end{figure*}

\section{Incidence of neutral outflows -- Mg {\small I} b - Na {\small I} D diagnostic}\label{neutraloutflows}

The Na {\small I} D$\lambda\lambda$5890,5896 absorption features can be used to directly probe the neutral phase of outflowing gas \citep[e.g., ][]{Rupke2005a,Villar2014}. Figure \ref{ppxf} shows a spectrum with a strong sodium line, well reproduced by the pPXF procedure, i.e. through stellar absorption. This is due to the fact that both stellar and interstellar absorption could be at the origin of the line. Therefore, in order to study the neutral outflowing gas component we need to discriminate between the two contributions. A simple diagnostic used to unveil the presence of neutral gas in the ISM is by comparing the EW of Na {\small I} D with that of Mg  {\small I} b triplet at $5167$, $5173$ and $5184\AA$, which is of pure stellar origin. Similar mechanisms by which sodium and magnesium are created in stars explain the simple relation EW(Na {\small I} D) $ =0.75\times$ EW(Mg {\small I} b) (\citealt{Heckman2000,Rupke2005a,Villar2014}); any deviation from that to higher EW(Na {\small I} D) is therefore interpreted as due to the presence of interstellar absorption in the Na {\small I} D.

For the sources in our sample we derive a first order estimate for EW(Mg {\small I} b) from the best-fit pPXF profile. The EW of Na {\small I} D is instead derived modelling the spectra around the absorption feature. He {\small I}$\lambda$5876 emission and Na {\small I} D absorption are fitted constraining the central wavelengths and FWHMs using the simultaneous fit results. The diagnostic comparing Mg {\small I}  b and Na {\small I}  D EWs is shown in Fig. \ref{NaIDMgIb} (left). The majority of AGNs shows Na {\small I} D dominated by stellar contribution (i.e., they lie on the above mentioned relation, shown in the figure as a dashed line), while only a small fraction of objects is located well below the correlation. 
When investigating separately the sub-samples with and without ionised [O {\small III}] outflows (blue and black symbols in the figure, respectively), we do not find any difference in their location on the above mentioned diagnostic.

Rupke et al. (2005a) found that the majority (80\%) of the galaxies below the relation EW(Na {\small I} D) $=3\ \times$ EW(Mg {\small I} b) have neutral outflows and proposed that objects in this region underway a starburst-driven outflow episode. 
The solid line in the diagram shows the relation reported by \citet{Rupke2005a}. Despite the high fraction of outflows found in ionized gas, this plot shows only one target below the Rupke et al. relation. This source, Mkn 848, is the only one for which we detect
a neutral outflow, according to this diagnostic diagram.

In Fig. \ref{NaIDMgIb} (right panels) we show the spectra around the Mg {\small I} b and Na {\small I} D absorption features for two sources with enhanced EW(Na {\small I} D). We note that while the magnesium profile is well fitted by pPXF procedure (orange curves), the Na {\small I} D absorption is strongly underestimated. In panels [II] and [IV] we show our fitting decomposition of the He {\small I} - Na {\small I} D system after removing the stellar continuum. Mkn 848 exhibits a broad absorption profile with a V$_{max}\approx 1250$ km/s, even higher than the maximum velocity associated with the ionized outflowing gas ($\approx 900$ km/s). For this source, the presence of neutral outflow on kpc scale, with a substantial mass rate ($\dot M_{out}^{Na ID}\approx 100$ M$_\odot$/yr) has been proved by spatially resolved spectroscopic analysis (\citealt{Rupke2013b}). 

The almost complete absence of neutral outflows in our sample is consistent with the results found by \citet{Villar2014}, who reported the lack of atomic outflows in a sample of 21 SDSS type 2 QSOs with detected ionized outflows. 
Instead, the fraction of outflows in Na {\small I} D increases up to 45\% in Seyfert ULIRGs \citep{Rupke2005a} and in sources exhibiting both SF and AGN activity \citep{Rupke2005b,Sarzi2016}. 
In the context of evolutionary scenario, these results suggest that the atomic outflows may be associated with earlier stages of the feedback phase, when the SF has not yet been completely inhibited and the AGN is still buried in dust enshrouded environments. This argument may explain the low incidence of atomic outflows and, on the whole, the low contribution of ISM absorption in our sample (see Fig. \ref{NaIDMgIb}, left), which is characterized by a large number of unobscured type 1 AGNs and, in general, by sources with high level of ionization (log[O {\small III}]/H$\beta\sim1$; fig. \ref{bpt}) as per their selection. 
The X-ray/SDSS target with atomic (and ionized) outflow signatures, Mkn 848, is a luminous infrared galaxy at z $=0. 04$ with high SFR (120 M$_\odot$/yr) associated with an ongoing merger (\citealt{Rupke2013b}), sharing therefore the same characteristics of other sources with unveiled neutral outflows. 
For example, Mkn 848 shares the same properties of other local ULIRG-QSO systems, such as Mrk 231, in which both cold (neutral and molecular) and warm outflows have been revealed in its host galaxy (see \citealt{Feruglio2015} and references therein). Furthermore, it is similar to other high-z sources analysed by our group (\citealt{Perna2015a,Cresci2015}): our analysis for the two quasars at z $\sim1.5$, XID 2028 and XID 5321 actually showed  concomitant intense star formation activity (SFR $\approx 250$ M$_\odot$/yr) and outflowing processes involving both atomic and ionized gas.

\section{X-ray loudness}\label{sdssloudness}

In this section we derive the X-ray bolometric corrections ($k_{bol}$) to test the X-ray loudness role in the outflow phenomena. We derive the bolometric corrections computing the ratio between the intrinsic X-ray luminosity and $L_{bol}$, as derived in Sect. \ref{nuclearSDSS}. We compare our results with the relation found by \citet{Lusso2012} and derived studying the X-ray selected AGN population in the COSMOS field. 

Figure \ref{kbolcorr} (left) shows the derived $k_{bol}$ against the AGN bolometric luminosities for our X-ray/SDSS sample (squared symbols). In the figure, large circles mark BL AGNs. We note that the latter sub-sample is likely associated with more reliable $L_{bol}$ estimates, being the values relative to the low-luminosity population corrected for host contamination (see Sect. \ref{nuclearSDSS}). Therefore, we focus the analysis on the BL sub-sample (which, however, include 80\% of the entire sample). For these sources, we differentiate between targets with kinematic indication of outflows in the [O {\small III}] line (with $V_{max}>650$ km/s), colour-coded for increasing velocities, and those with narrow [O {\small III}] profiles (empty circles).    

From a visual inspection, it appears that the X-ray/SDSS sources are broadly distributed over the entire plane, regardless the presence/absence of signatures of outflows in their ionized gas component. To simplify the visualization, we also show in Fig. \ref{kbolcorr} (right) the source distribution normalizing (in log space) the $k_{bol}$ with the value expected from the \citet{Lusso2012} relation ($k_{bol-L12}$; black curve in the left panel), considering only the luminosity range in which the relation has been calibrated (i.e., log($L_{bol}/L_\odot $) in the range $10.8-13.5$). The histogram shows that the entire sample of BL AGNs (grey curve) is actually following the expected relation for X-ray selected AGNs, displaying a Gaussian distribution. 
Most important, the distributions of sources with/without outflows are similar, i.e. we do not report any excess in the number of sources with outflows at negative log($k_{bol}/k_{bol-L12}$). This result therefore seems to disprove the idea of a X-ray loudness associated with the feedback phase. 

The $k_{bol}$ correction derived for the sources presented in our previous works (\citealt{Brusa2015,Perna2015a,Brusa2016}), and for which we were able to derive well constrained bolometric luminosities (from SED fitting technique) and intrinsic X-ray emission (fitting XMM spectra) are very solid. Although the X-ray/SDSS $k_{bol}$ corrections are probably less reliable,  the almost symmetric distribution with respect to the expected relation found by \citet{Lusso2012} suggests that we are actually able to derive good estimates for the X-ray bolometric corrections\footnote{
Figure \ref{kbolcorr} shows that X-ray/SDSS sources are characterized by more scattered distribution if compared with those of \citet{Lusso2012}. This is reasonably due to our approach in deriving bolometric luminosities, which requires an empirical relation based on monochromatic luminosity or, for fainter objects, two empirical relations (see Sect. \ref{nuclearSDSS}). At lower luminosities (log($L_{bol}/L_\odot) <11.5$) in the figure), $L_{bol}$ and $k_{bol}$ errors are of the order of a factor of $\sim$ 2; at higher luminosities, they are of the order of 10-30\%. 
}. It is possible therefore that the lower $k_{bol}$ corrections found in the high-z sample are due to the fact that the sources of the two samples trace two different evolutionary phases, with distinct properties.

The different incidence of atomic outflows between the X-ray/SDSS sample dominated by unobscured AGNs and the high-z X-ray (and optical) obscured QSOs sample points in the same direction.    
In such a context, the X-ray/optical obscured AGNs studied in the previous works are associated with the initial stages of the feedback, when atomic gas is still present in the ISM and outflow processes involve all the gas components (e.g., XID 2028 and XID 5321), while the BL AGNs are tracing later stages, in which the line of sight has been cleaned and the cold components have been heated or exhausted. 
We note, however, that our suggestions are still based of small samples of obscured QSOs and that further investigation is needed to understand if the observed physical conditions at larger scales involving neutral ISM gas are somehow related with the accretion processes responsible of the X-ray emission.

\section{Summary}\label{sdssdiscussion}

We analysed SDSS optical and X-ray spectra of a large sample of 563 AGN at  z $<0.8$, comprising type 1 (362), type 1.9 (77) and type 2 (124) sources. We combined ionized emission line and neutral absorption feature information modelled  through multicomponent simultaneous fitting, non-parametric measurement and pPXF analysis, in order to derive physical (e.g., ionized levels) and kinematic conditions of both warm and cold gas components of the ISM. The main results are reported below.

Analysing the [O {\small III}] line profiles, we derived the outflow incidence
in the warm phase of the NLR: we found that  $\approx$ 40\% of
AGNs exhibit signatures of outflows. Such fraction is strongly
dependent on the AGN power: our analysis suggest a clear positive
correlation between the outflow velocity and the AGN power
traced by the [O {\small III}] luminosity. Such trend has been already reported in the literature
(e.g., \citealt{Bae2014}) and also related with similar correlations
between radio luminosity and [O {\small III}] width (e.g., \citealt{Mullaney2013,Zakamska2016,Woo2016}). However, the observed correlation between radio
luminosity and stellar mass (and stellar velocity dispersion) may
easily explain the trend with the [O {\small III}] widths, resulting in a
challenging interpretation (see, e.g., the discussion in \citealt{Woo2016,Zakamska2016}).
We show instead, for the first time on a large sample ($\sim $ 550 AGNs), that a well defined positive trend of outflow velocity is observed with the unobscured $2-10$ keV X-ray luminosity, a tracer unambiguously associated with AGN activity, and it holds over 5 order of magnitudes. 

Broad absorption line (BAL) quasars are generally associated with AGN-driven winds expelling UV/Optical absorbing material at velocities such high as $\sim 40000$ km/s (e.g., \citealt{Dunn2010,Kaastra2014}). We note that this class of objects has been recently associated with intrinsically weak X-ray sources (e.g., \citealt{Luo2014}). The simplest explanation for the discrepancy between these results and the positive correlation we found between [O {\small III}] outflow velocity and X-ray luminosity is that UV/optical absorbing ejected material and ionized outflows may be related to different physical processes happening at different spatial scales (see, e.g., \citealt{Fiore2017}).

The sodium absorption system at $\lambda\lambda$5890,5896 has been analysed to infer the presence of atomic outflowing gas in the ISM. We found signatures of atomic outflow in only one target and derived an incidence for the atomic outflows much lower ($<1\%$) than the one obtained for the ionized counterpart.

We derived the X-ray bolometric correction and proved that the X-ray/SDSS sample characterized by the presence of ionized outflows do not show any deviation from the typical behaviour of the population of X-ray selected AGNs. This result may rule out the proposed role of X-ray emission in the feedback phase (Sect. \ref{sdssloudness}; \citealt{Brusa2016}). However, we note that our selection criteria allowed the collection of a sample mostly dominated by BL AGNs which, in the framework of the evolutionary scenario, may be associated with a different stage of the blow-out phase (e.g., \citealt{Hopkins2008}). 
We therefore suggested a scenario which explain the different evidences of neutral/ionized outflows and nuclear properties (i.e. X-ray activity), invoking two different evolutionary stages within the blow-out phase: an initial X-ray/optical obscured stage, in which the atomic gas is still present in the ISM and the outflow processes involve all the gas components, and a later stage associated with unobscured AGNs, which line of sight has been cleaned and the cold components have been heated or exhausted.

\vspace{2cm}

{\small 
{\it Acknowledgments:} MP, GL and MB acknowledge support from the FP7 Career Integration Grant ``eEASy'' (``SMBH evolution through cosmic time: from current surveys to eROSITA-Euclid AGN Synergies'', CIG 321913).
Support for this publication was provided by the Italian National Institute for Astrophysics (INAF) through 
PRIN-INAF-2014  (``Windy  Black  Holes
combing  galaxy  evolution''). We thank the anonymous referee for his/her constructive comments to the paper.

    Funding for the Sloan Digital Sky Survey (SDSS) has been provided by the Alfred P. Sloan Foundation, the Participating Institutions, the National Aeronautics and Space Administration, the National Science Foundation, the U.S. Department of Energy, the Japanese Monbukagakusho, and the Max Planck Society. The SDSS Web site is \url{http://www.sdss.org/}.

    The SDSS is managed by the Astrophysical Research Consortium (ARC) for the Participating Institutions. The Participating Institutions are The University of Chicago, Fermilab, the Institute for Advanced Study, the Japan Participation Group, The Johns Hopkins University, Los Alamos National Laboratory, the Max-Planck-Institute for Astronomy (MPIA), the Max-Planck-Institute for Astrophysics (MPA), New Mexico State University, University of Pittsburgh, Princeton University, the United States Naval Observatory, and the University of Washington.
}

\begin{appendix}
\section{\\Fe {\small II} emission line template}\label{AppendixA}
We use theoretical model templates of \citet{Kovacevic2010} to reproduce Fe {\small II} emission.
The authors constructed a Fe {\small II} template consisting of tens of line components, identified as the strongest within $4000-5500\AA$ range. Many of them are sorted into five line groups according to the lower term of their atomic transition (P, F, S, G and H, see their fig. 1). Relative intensities of each Fe {\small II} Gaussian line within a single group have been calculated studying the transition probabilities. A sixth group, whose relative intensities have been obtained on the basis of their best-fit of the well-studied I Zw 1 spectrum, was added to reproduce the overall Fe {\small II} profile. 
These templates assume seven free fitting parameters: the shift relative to the systemic redshift and the FWHM of the Fe {\small II} lines, which are assumed to be the same for all the lines of the template, and the intensities of the Fe {\small II} lines within the line groups. 
Iron emission strongly cover the entire wavelength range of H$\beta$ and [OIII]. As a result, a correct modelling is needed to fully characterize the Balmer profile, and in particular its FWHM, from which crucial information can be derived, and the double ionised oxygen emission. One of the S group line (see \citealt{Kovacevic2010}, fig. 1) may be responsible for a red excess in the [O {\small III}] line, leading to a wrong determination of the kinematic conditions in the ionised gas. The fact that the amplitude of this line is linked to the intensity of the entire S group (and, in particular, to the nearest line in the red Fe  {\small II} bump) gives and important contribution to the deconvolution between Fe {\small II} and reddest emission from [O {\small III}]. As an example, we show in Fig. \ref{feiisubtraction} the fitting results we obtain for a couple of SDSS sources, in which prominent red wings in the [O {\small III}] line are detected.

\begin{figure}[b]
\centering
\includegraphics[width=9.4 cm,trim=0 130 0 260,clip]{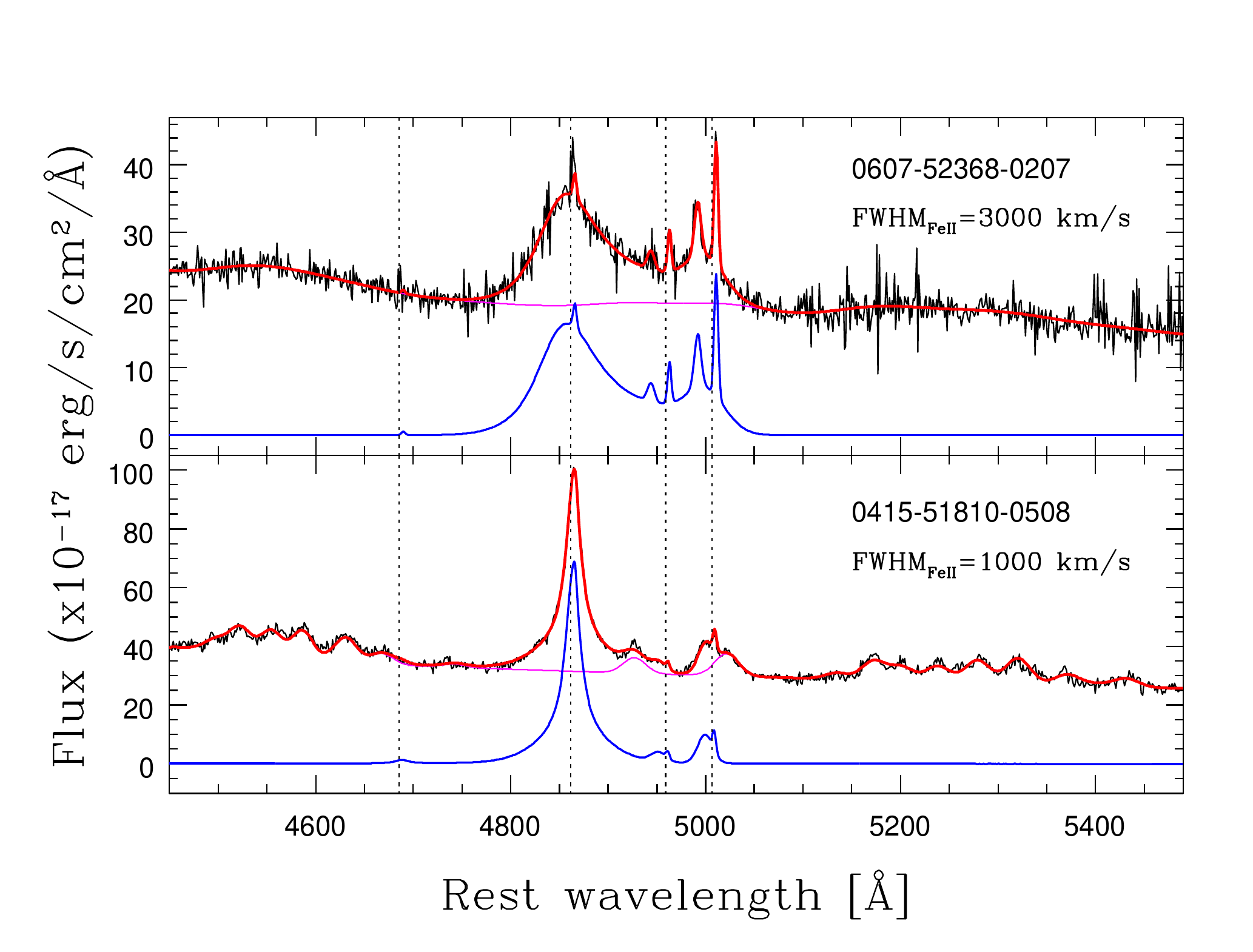}
\caption{\small Multicomponent simultaneous best-fit results (red curves) for two X-ray/SDSS sources showing iron emission and double peaked [O {\small III}] profiles. Black profiles represent rest-frame SDSS spectra. Blue and magenta curves mark the best-fit results for the [O {\small III}]-H$\beta$ emission and for the Fe {\small II} lines respectively. For visual inspection, the iron profile is added to the power-law continuum. The figure shows the deconvolution between Fe {\small II} and the reddest emission from [O {\small III}]. We labelled in the panels the MJD, the plate and the fibre numbers which uniquely identifies the SDSS spectra and the FWHM required to fit the iron emission.}
\label{feiisubtraction}
\end{figure}

\vspace{3cm}

\section{\\ $[$O {\small III}$]$  velocity - luminosity trend. Supporting evidences from literature}\label{AppendixC}

In Fig. \ref{W80lumlit} we compare our results with those obtained studying other large samples of SDSS type 2 AGNs in the same redshift range and for which non-parametric $W80$ velocity estimates are available\footnote{\url{http://zakamska.johnshopkins.edu/data.htm}}. \citet[$\sim 2900$]{Yuan2016} and \citet[$\sim 550$]{Reyes2008} sources are shown with grey and pink symbols respectively. Their luminosity range ($\sim 10^{42}-10^{43}$ erg/s) is much shorter than the one covered by our X-ray/SDSS sources, and do not permit a real comparison of velocity - luminosity trends; however, when we derive the median values in luminosity bins for our sources (red/yellow large symbols) and for the total sample obtained combining our sources with theirs (blue/cyan large symbols), we do not observe any deviation at higher luminosities ($\gtrsim 10^{42}$ erg/s). The large number ($\sim 4000$) of sources, together with the use of median instead of average values in luminosity bins (less sensitive to too much deviated velocity estimates) supports the results we have shown in Fig. \ref{LvsV}.

\begin{figure}[]
\centering
\includegraphics[width=9.4 cm,trim=0 130 0 140,clip]{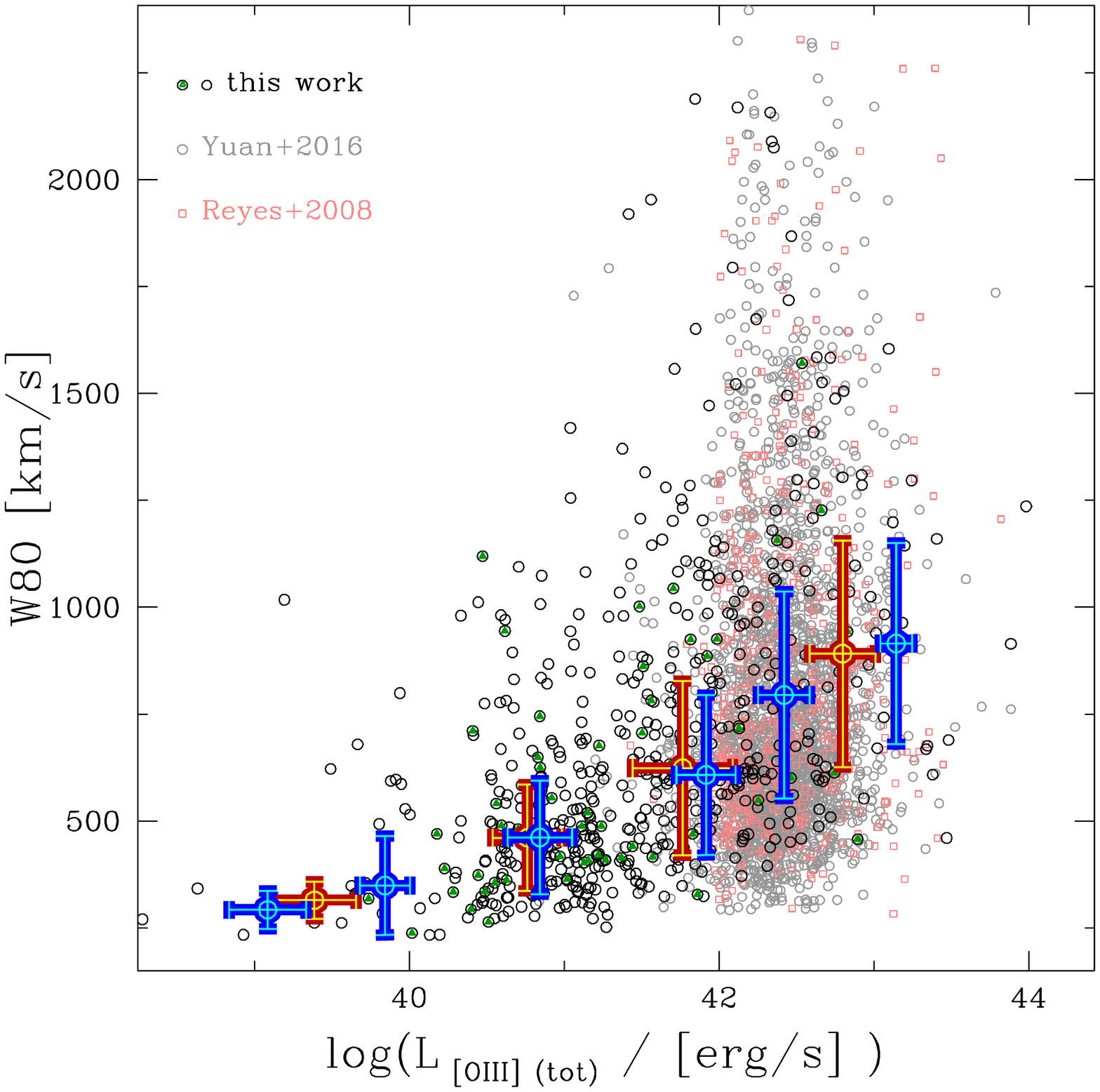}
\caption{\small 
Non-parametric velocity $W80$ estimator as a function of [O {\small III}] luminosity. Black and green/black symbols refer to the same X-ray/SDSS sources shown in Fig \ref{LvsV}; grey circles and pink squares shows type 2 AGNs from \cite{Yuan2016} and \cite{Reyes2008}. Large symbols represent median velocities in bins of luminosity for the X-ray/SDSS sample (red/yellow) and for all sources in the figure (blue/cyan); bars indicate semi-interquartile ranges. Yuan et al. and Reyes et al. luminosities have been re-computed consistently with the cosmological parameters displayed in the Introduction Section.
}
\label{W80lumlit}
\end{figure}

\section{Signal to noise ratio and NC/OC flux ratio investigation}\label{AC3}

In order to prove that the correlations displayed in Sect. \ref{sdssincidence} are not due to detection biases, we investigate whether the absence of fast outflows at low luminosities and of slow outflows at highest luminosities are related to particular OC/NC flux ratios and/or S/N of the OC Gaussians.

Figure \ref{correlationsAC}, panel $a$, shows the distribution of $f_{OC}/f_{NC}$ versus $V_{max}$ (in log-space) for the sub-sample modelled with NC+OC Gaussians.
A poor positive correlation (SR $=0.18$, with null hypothesis $\sim10^{-5}$) is observed. 
This correlation is expected, as non-parametric velocity estimators are computed from total [O {\small III}] flux percentiles (Sect. \ref{sdssanalisi}) and, when broad prominent wings are present, both relative OC/NC flux contributions and $V_{max}$ are expected to increase. The correlation is inconspicuous because of the dependence of maximum velocity non the widths of Gaussian components, which are not taken into account here.

Figure \ref{correlationsAC}, panel $e$, shows the distribution of OC signal to noise ratio versus $V_{max}$ (log-space). Also in this case, we observe a poor (negative) correlation (SR $=-0.26$, $P\approx 10^{-6}$). The correlation suggests negligible effects of bias selections: naively, highest velocities could be uniquely associated with objects with prominent (i.e., well detected), extended wings. If it is the case, a positive correlation should be observed in our SN$_{OC}$ - $V_{max}$ distribution. On the other hand, it is possible that highest velocities could be wrongly derived when [O {\small III}] wings are characterized by low S/N. We mitigated as best as possible this effect with simultaneous fit modelling (see Sect. \ref{sdssanalisi}); the poor correlation prove the capabilities of this technique.

Panels $b$, $c$, $f$ and $g$ display the distributions of OC/NC flux ratios and S/N$_{OC}$ with respect to [O {\small III}] and X-ray luminosities. In these cases, we do not observe any relevant correlation (SR $\sim -0.1 \div 0.1$, $P\sim 1\div50\%$). We can therefore reasonably exclude any contribution of low S/N or peculiar OC/NC to the correlations shown in Sect. \ref{sdssincidence} between outflow velocity and AGN luminosity.

Finally, panel $d$  and $h$ show the distributions of NC/OC and S/N$_{OC}$. We note that the OC contribution to the total oxygen flux is, on average, of the same order of magnitude of NC flux. The S/N$_{OC}$ distribution shows that the majority of our sample modelled with NC+OC Gaussians also presents significant OC emission (S/N$>3$). Only $\sim20$ sources are below this threshold. We do not exclude any of these as the presence of OC Gaussians has been proved also in other emission lines in the H$\beta$ and H$\alpha$ regions. We note also that only $\sim$15 targets enter in the determination of the fraction of AGNs with outflows ($V_{max}>665$ km/s; Sect. \ref{sdssincidence}). Therefore, their contribution do not determine significant variation in our results.

\begin{figure}[]
\centering
\includegraphics[width=9.4 cm,trim=0 150 0 90,clip]{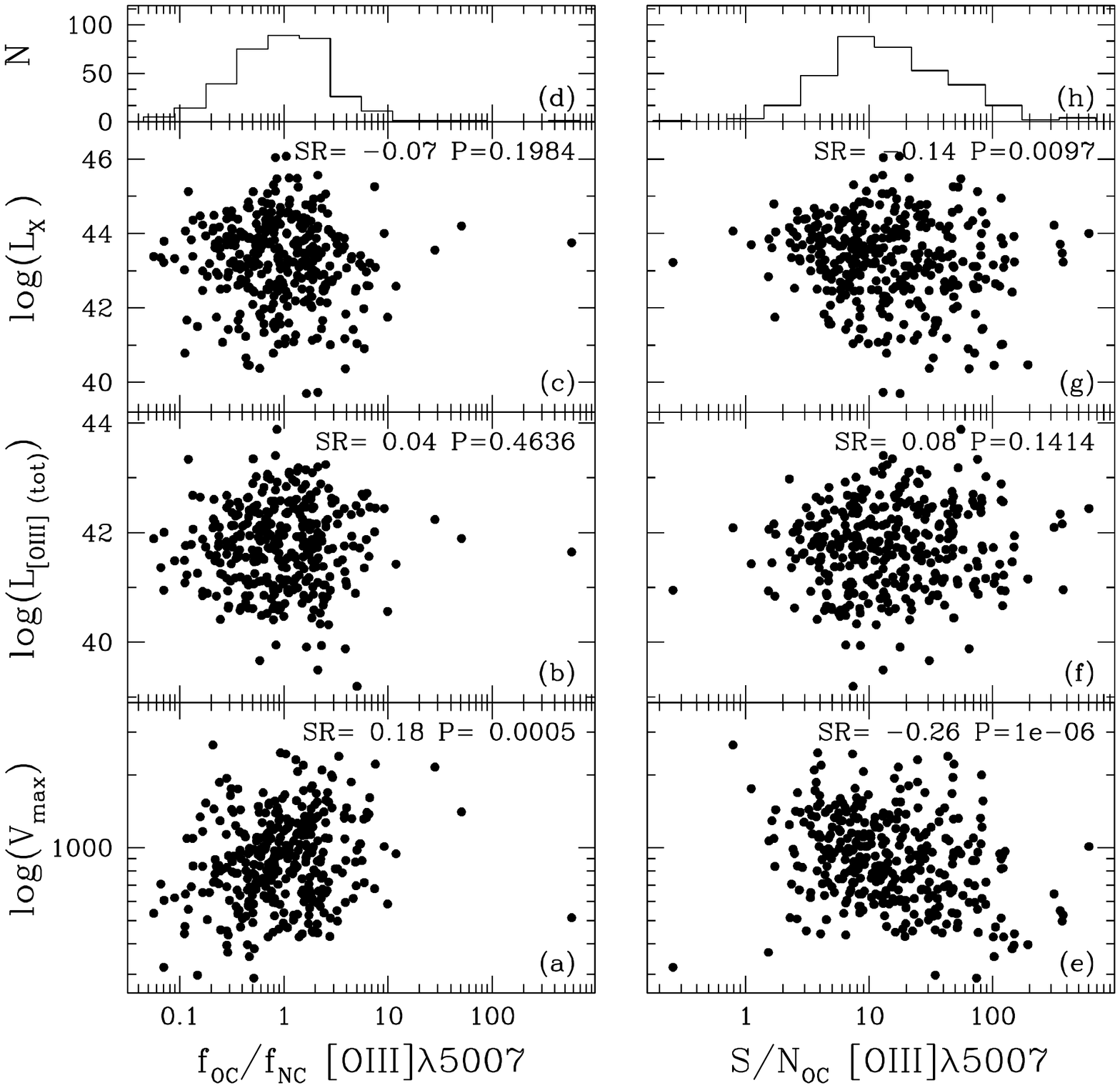}
\caption{\small 
OC/NC flux ratios and [O III]$\lambda$5007 OC signal to noise (S/N$_{OC}$) as a function of $V_{nax}$, [O III] and X-ray luminosities for the sub-sample modelled with NC+OC Gaussians. Spearman rank coefficients and null hypothesis probabilities are also labelled in each panel. The panels in the first row display the distributions of OC/NC ratios and S/N$_{OC}$. 
}
\label{correlationsAC}
\end{figure}

\end{appendix}

\end{document}